\documentclass[12pt,a4paper]{article}
\usepackage{epsfig}
\usepackage{amsfonts}
\usepackage{amssymb}

\begin{document}
\begin{flushright}ITEP-TH-12/12
\end{flushright}
\begin{center}

\vspace{1cm}

{\bf \large ON NMHV FORM FACTORS IN $\mathcal{N}=4$ SYM THEORY FROM
GENERALIZED UNITARITY} \vspace{2cm}

{\bf \large L. V. Bork}\vspace{0.5cm}

{\it Institute for Theoretical and Experimental Physics, Moscow,
Russia.}\vspace{1cm}

\abstract{In this paper a supersymmetric version of a generalized
unitarity cut method in application to MHV and NMHV for form factors
of operators from the $\mathcal{N}=4$ SYM stress-tensor current
supermultiplet $T^{AB}$ at one loop is discussed. The explicit
answers for 3 and 4 point NMHV form factors at tree and one loop
level are obtained. The general structure of n-point NMHV form
factor at one loop is discussed as well as the relation between form
factor with super momentum equal to zero and the logarithmic
derivative of the superamplitude with respect to the coupling
constant.}
\end{center}


\newpage

\tableofcontents{}\vspace{0.5cm}

\renewcommand{\theequation}{\thesection.\arabic{equation}}
\section{Introduction}\label{s1}
Much attention in the past decade has been  paid to the study of the
scattering amplitudes (the $S$-matrix) in four dimensional gauge
theories, especially in the planar limit of $\mathcal{N}=4$ SYM
theory.

It is believed that the hidden symmetries of the $\mathcal{N}=4$ SYM
theory which are responsible for its integrability properties will
completely fix the structure of the amplitudes. The hints that the
$S$-matrix for the $\mathcal{N}=4$ SYM theory can be fixed by some
underlying integrable structure were found at weak
\cite{BeisertYangianRev,BeisertYangianAmpl,Gorsky:2009ew} and strong
\cite{Gorsky:2009ew,Alday:2010vh} coupling regimes.

There is another class of objects of interest in the $\mathcal{N}=4$
SYM theory which resembles the amplitudes -- the form factors which
are the matrix elements of the form
\begin{equation}
\langle p_1^{\lambda_1}, \ldots,
p_n^{\lambda_n}|\mathcal{O}|0\rangle,
\end{equation}
where $\mathcal{O}$ is some gauge invariant operator which acts on
the vacuum and produces some state $|p_1^{\lambda_1}, \ldots,
p_n^{\lambda_n} \rangle$ with momenta $p_1, \ldots, p_n$ and
helicities $\lambda_1, \ldots, \lambda_n$\footnote{Note that
scattering amplitudes in ''all ingoing" notation can schematically
be written as $\langle p_1^{\lambda_1}, \ldots, p_n^{\lambda_n}|0
\rangle$. }. The $S$-matrix operator is assumed in both cases. One
can think about this object as an amplitude of the proses where
classical current couples through gauge invariant operator produces
quantum state. The example of such process is $\gamma^{*}\to
\mbox{Jet's}$ in perturbative QCD \cite{Kazakov:1987jk} 
(see also \cite{HuberNew1,HuberNew2} for recent results) where we
take into account all orders in $\alpha_s$ but the first order in
$\alpha_{em}$. The amplitude of such process is given by the matrix
element of the following form:
$$
\langle p_1^{\lambda_1} \ldots
p_n^{\lambda_n}|j^{QCD}_{em}|0\rangle,$$ where $j^{QCD}_{em}$ is the
QCD quark electromagnetic current operator.

The two-point form factors in $\mathcal{N}=4$ SYM were studied long time ago in
\cite{vanNeerven:1985ja} and recently in \cite{Henn3loops}. Using the $\mathcal{N}=3$
superfield formalism the form factors of none gauge invariant
operators (off-shell currents) at tree level were derived in
\cite{Perturbiner}. Recently, the strong coupling limit of  form
factors has been studied in \cite{Maldacena:2010kp} and the weak
coupling regime in
\cite{BKV_FormFN=1,FormFactorMHV_component_Brandhuber,BKV_SuperForm,HarmonyofFF,FF_MHV_3_2loop}.
Also different regularizations for form factors were discussed in\cite{HennReg}.
The motivations for the systematic study of form factors in
$\mathcal{N}=4$ SYM are
\begin{itemize}
\item it might help to understand
the symmetry properties of the amplitudes
\cite{BeisertYangianRev,BeisertYangianAmpl}. It is believed that the
symmetries completely fix the amplitudes of the $\mathcal{N}=4$ SYM
theory and  it is interesting to see whether they fix/restrict the
form factors as well;

\item the form factors are the intermediate objects between
the fully on-shell quantities such as the amplitudes and the fully
off-shell quantities such as the correlation functions (which are
one of the central objects in AdS/CFT). Since the powerful
computational methods have appeared recently for the amplitudes in
$\mathcal{N}=4$ SYM ( see, for example,
\cite{BCF,N=4onshellmethods0,N=4onshellmethods1,N=4onshellmethods2} and
\cite{AHamed_Integrands, NMHV_2loop_Dixon}), it would be desirable
to have some analog  of them for the correlation functions
\cite{Raju:2011mp}. The understanding of the structure of form
factors and the development of computational methods  might shed
light on the correlation functions;

\item also, it might be useful for understanding of
the relation between the conventional description of the gauge
theory in terms of local operators and its (possible) description in
terms of Wilson loops. The latter fact is the so-called
amplitude/Wilson loop duality which originated for the case of
$\mathcal{N}=4$ SYM in \cite{minSurface4point,dualKorch,dual}. This
duality was intensively studied in the weak and strong coupling
regimes and tested in different cases, and its generalizations to
the non-MHV amplitudes were proposed in \cite{Huot}. Moreover such
dual description for amplitudes of $\mathcal{N}=4$ SYM together with
the developments of OPE technique for Wilson loops
\cite{WLoopOPE1,WLoopOPE2} led to the formulation of the equation
\cite{HuotEquation} which in principle should define the whole
$\mathcal{N}=4$ SYM S-matrix for any value of coupling constant. 
Note that similar equation for the
amplitudes may be derived from "twistor space" (see for example \cite{Twistors1}) 
point of view \cite{Twistors2DescentEquation}. It
is interesting to investigate whether such dual description for form
factors in $\mathcal{N}=4$ SYM exists and if it is possible to
formulate this equation for form factors. 
\end{itemize}

To make progress in the above-mentioned directions, the perturbative
computations at several first orders of perturbative theory are
likely required.

The aim of this paper is to continue investigations of the form
factors \cite{BKV_SuperForm,HarmonyofFF} of the operators from stress tensor
supermultiplet in $\mathcal{N}=4$ SYM. We will use the
formulation of form factors in $\mathcal{N}=4$ in
on-shell momentum superspace developed in
\cite{BKV_SuperForm}, which will allow us to consider form
factors with different types of particles in $\langle
p_1^{\lambda_1}, \ldots, p_n^{\lambda_n}|$ external state in
$\mathcal{N}=4$ covariant manner. We will use the generalized
unitarity technique to study the structure of NMHV sector at one
loop. First we are going to discuss how the generalized unitarity
technique works for form factors for MHV sector at one loop in
$\mathcal{N}=4$ on-shell momentum superspace. Then we will continue
with the NMHV sector. We will perform explicit computations of 3 and
4 point NMHV form factors at one loop and will discuss the structure
of the general n point situation. We make a brief comment on the
relation between form factors with operator insertion with zero
momentum and amplitudes.

\section{Amplitudes and form factors in on-shell momentum superspace}\label{s2}

\subsection{Super form factors of the chiral truncation of $\mathcal{N}=4$ stress tensor supermultiplet}
To describe the $\mathcal{N}=4$ SYM stress tensor supermultiplet it
is convenient to use standard $\mathcal{N}=4$ coordinate superspace
\begin{equation}
\mbox{$\mathcal{N}=4$ coordinate
superspace}=\{x^{\alpha\dot{\alpha}},~\theta^A_{\alpha},~\bar{\theta}_{A\dot{\alpha}}\},
\end{equation}
where $x_{\alpha\dot{\alpha}}$ are bosonic coordinates and
$\theta$'s, which are $SU(4)_R$ vectors and Lorentz
$SL(2,\mathbb{C})$ spinors, are fermionic ones. The $\mathcal{N}=4$
supermultiplet of fields (containing $\phi^{AB}$ scalars
(anti-symmetric in $SU(4)_R$ indices $AB$), $\psi^A_{\alpha},
\bar{\psi}^A_{\dot{\alpha}}$ fermions and $F^{\mu\nu}$-- the gauge
field strength tensor, all in the adjoint representation of
$SU(N_c)$ gauge group) is realized in the $\mathcal{N}=4$ coordinate
superspace as the constrained superfield $
W^{AB}(x,\theta,\bar{\theta})$ with the lowest component
$~W^{AB}(x,0,0)=\phi^{AB}(x)$. $W^{AB}$ in general is not a chiral
object and satisfies several constraints
\cite{SuperCor1,BKV_SuperForm}: a self-duality constraint
\begin{equation}
W^{AB}(x,\theta,\bar{\theta})
=\overline{W_{AB}}(x,\theta,\bar{\theta})=
\frac{1}{2}\epsilon^{ABCD}W_{CD}(x,\theta,\bar{\theta}),
\end{equation}
which implies $\phi^{AB}=\overline{\phi_{AB}}=\frac 12
\epsilon^{ABCD} \phi_{CD}$ and two additional
constraints\footnote{$[\ast,\star]$ denotes antisymmetrization in
indices, while $(\ast,\star)$ denotes symmetrization in indices.}
\begin{eqnarray}\label{PartialChirality} &&
D_C^{\alpha}W^{AB}(x,\theta,\bar{\theta}) = -\frac{2}{3}\delta^{[A}_CD_L^{\alpha}W^{B]L}(x,\theta,\bar{\theta}), \nonumber\\
&&\bar{D}^{\dot{\alpha}(C}W^{A)B}(x,\theta,\bar{\theta}) = 0,
\end{eqnarray}
where $D^A_{\alpha}$ is a standard coordinate superspace
derivative\footnote{which is
$D^A_{\alpha}=\partial/\partial\theta_{A}^{\alpha}
+i\bar{\theta}^{A\dot{\alpha}}\partial/\partial
x^{\alpha\dot{\alpha}}$.}. Note that in this formulation full
$\mathcal{N}=4$ supermultiplet of fields is on-shell in the sense
that the algebra of the generators
$Q^{A}_{\alpha},\bar{Q}_{B\dot{\alpha}}$ of the supersymmetric
transformation of the fields in this supermultiplet is closed only
if the fields obey their equations of motion. $\mathcal{N}=4$
SYM stress tensor supermultiplet $T^{AB}$ is given then by
\begin{equation}
 T^{AB}=\mbox{Tr}\left(W^{AB}W^{AB}\right).
\end{equation}

We will consider in this article the chiral truncation of
$\mathcal{N}=4$ SYM stress tensor supermultiplet (which contains
only self-dual part of full multiplet) rather than the
supermultiplet itself \cite{SuperCor1,BKV_SuperForm,HarmonyofFF}. 
The main reason for this
is that the chiral truncation has the off-shell description in terms
of superfields on $\mathcal{N}=4$ superspace i.e. the component
fields in such truncated multiplet are arbitrary and the chiral part
of the algebra of supersymmetric transformations of the component
fields can be still closed without any constraints on the component
fields \cite{SuperCor1}. Note that the off-shell description for the full
$\mathcal{N}=4$ supermultiplet in any superspace is unknown.

To describe this truncated supermultiplet 
\cite{SuperCor1} one has to break $SU(4)_R$
group into two $SU(2)$ and $U(1)$
\begin{eqnarray}\label{projection1}
SU(4)_R &\rightarrow& SU(2)\times SU(2)'\times U(1),
\end{eqnarray}
so that the index $A$ of $R$-symmetry group $SU(4)_R$ splits into
\begin{eqnarray}\label{prrojection2}
A&\rightarrow&(+a|-a'),
\end{eqnarray}
where $+a$ and $-a'$ correspond to two copies of $SU(2)$ and $\pm$
correspond to the $U(1)$ charge. We will not write the  $U(1)$
factor explicitly hereafter, and will use a notation
\begin{eqnarray}\label{prrojection2}
(+a|-a') \equiv (a|\dot{a}).
\end{eqnarray}
After that one has to take the particular ($ab$) projection of
$W^{AB}$ that depends on half of the Grassmann coordinates
\cite{BKV_SuperForm} (this can be seen from the
(\ref{PartialChirality})):
$W^{ab}(x,\theta^{c},\bar{\theta}_{\dot{c}})$. The truncated stress
tensor supermultiplet is then given by
\begin{equation}\label{Wchiral}
\mathcal{T}^{ab}=Tr\left(W^{ab}W^{ab}\right)\Big|_{\bar{\theta}=0}.
\end{equation}

To describe external states in $\mathcal{N}=4$ covariant manner it
is convenient to use $\mathcal{N}=4$ on-shell momentum superspace
\cite{DualConfInvForAmplitudesCorch}. This superspace is
parameterized in terms of $SL(2,\mathbb{C})$ spinors
$\lambda_{\alpha}, \tilde{\lambda}_{\dot{\alpha}}, \alpha,
\dot{\alpha}=1,2$ and Grassmannian coordinates $\eta^A, A=1,
\ldots,4$ which are Lorentz scalars and $SU(4)_R$ vectors
\begin{equation}
\mbox{On-shell $\mathcal{N}=4$ momentum superspace} =
\{\lambda_{\alpha}, \tilde{\lambda}_{\dot{\alpha}}, \eta^A\}.
\end{equation}
In this superspace the creation/annihilation operators
$$
\{g^-,~\Gamma^A,~\phi^{AB},~\bar{\Gamma}^A,~g^+\},
$$
of $\mathcal{N}=4$ supermultiplet, for the on-shell states which
are two physical polarizations of gluons $|g^-\rangle, |g^+\rangle$,
four fermions $|\Gamma^A\rangle$ with positive and four fermions
$|\bar{\Gamma}^A\rangle$ with negative helicity, and three complex
scalars $|\phi^{AB}\rangle$ (anti-symmetric in $SU(4)_R$ indices
$AB$ ) can be combined together into one $\mathcal{N}=4$ invariant
superstate ("superwave-function")
$|\Omega_{i}\rangle=\Omega_{i}|0\rangle$:
\begin{eqnarray}\label{superstate}
|\Omega_{i}\rangle=\left(g^+_i + \eta^A\Gamma_{i,A} +
\frac{1}{2!}\eta^A\eta^B \phi_{i,AB} +
\frac{1}{3!}\eta^A\eta^B\eta^C \varepsilon_{ABCD}\bar{\Gamma}^D_i +
\frac{1}{4!}\eta^A\eta^B\eta^C\eta^D \varepsilon_{ABCD}g^-_i\right)
|0\rangle,
\nonumber\\
\end{eqnarray}
where $i$ corresponds to the on-shell momentum
$p^i_{\alpha\dot{\alpha}} =
\lambda_{\alpha}^i\tilde{\lambda}_{\dot{\alpha}}^i$,  $p_i^2=0$
carried by the particle. The $n$ particle external state
$|\Omega_n\rangle$ is then given by
$|\Omega_n\rangle=\prod_{i=1}^n\Omega_i|0\rangle$.

The form factor $\mathcal{F}_n$ of the truncated stress tensor
supermultiplet for general n particle external state is then given
by:
\begin{eqnarray}
\mathcal{F}_n(\{\lambda,\tilde{\lambda},\eta\},q,\theta^{a})=
\langle\Omega_n|\mathcal{T}_{ab}(x,\theta^{a})|0\rangle,
\end{eqnarray}
where $\{\lambda,\tilde{\lambda},\eta\}$ is short notation for
$(\lambda_1,\tilde{\lambda}_1,\eta_1\ldots
\lambda_n,\tilde{\lambda}_n,\eta_n)$. Here we are considering colour
ordered object $\mathcal{F}_n$. The physical form factor
$\mathcal{F}_n^{phys.}$ in the planar limit\footnote{$g \rightarrow
0$ and $N_c \rightarrow \infty$ of $SU(N_c)$ gauge group so that
$\lambda=g^2N_c=$fixed.} should be obtained from $\mathcal{F}_n$ as:
\begin{eqnarray}
\mathcal{F}_n^{phys.}(\{\lambda,\tilde{\lambda},\eta\},q,\theta^{a})=(2\pi)^4g^{n-2}2^{n/2}\sum_{\sigma\in
S_n/Z_n}Tr(t^{a_{\sigma(1)}}\ldots
t^{a_{\sigma(n)}})\mathcal{F}_n(\sigma(\{\lambda,\tilde{\lambda},\eta\}),q,\theta^{a}),\nonumber\\
\end{eqnarray}
where the sum runs over all possible none-cyclic permutations
$\sigma$ of the set $\{\lambda,\tilde{\lambda},\eta\}$ and the trace
involves $SU(N_c)$ generators $t^a$ in the fundamental
representations. The normalization $Tr(t^at^b)=1/2$ is used.

Performing Fourier transform for bosonic coordinate
$x^{\alpha\dot{\alpha}}\rightarrow q_{\alpha\dot{\alpha}}$ and
taking into account that $\mathcal{F}_n$ is chiral, translationally
invariant and $\mathcal{T}^{ab}$ is $1/2$-BPS we see that
$\mathcal{F}_n$ should satisfy the following conditions
\cite{BKV_SuperForm}:
\begin{eqnarray}
P_{\alpha\dot{\alpha}}\mathcal{F}_n=Q^a_{\alpha}\mathcal{F}_n=Q^{\dot{a}}_{\alpha}\mathcal{F}_n=\bar{Q}_{a\dot{\alpha}}\mathcal{F}_n=0,
\end{eqnarray}
where generators of supersymmetry algebra
$(P_{\alpha\dot{\alpha}},Q^a_{\alpha},Q^{\dot{a}}_{\alpha},\bar{Q}_{a\dot{\alpha}},\bar{Q}_{\dot{a}\dot{\alpha}})$
acting on $\mathcal{F}_n$ are given by
\begin{eqnarray}\label{SUSYGenF}
\mbox{4 translations } P_{\alpha\dot{\alpha}}&=&
-\sum_{i=1}^n\lambda_{\alpha}^i\tilde{\lambda}_{\dot{\alpha}}^i+q_{\alpha\dot{\alpha}},\nonumber\\
\mbox{4 supercharges }
Q_{\alpha}^a&=&-\sum_{i=1}^n\lambda_{\alpha}^i\eta^{a}_i
+\frac{\partial}{\partial\theta^{\alpha}_a },\nonumber\\
\mbox{4 supercharges }
Q_{\alpha}^{\dot{a}}&=&-\sum_{i=1}^n\lambda_{\alpha}^i\eta^{\dot{a}}_i
+ \frac{\partial}{\partial\theta^{\alpha}_{\dot{a}} },\nonumber\\
\mbox{4 conjugated supercharges }
\bar{Q}_{a\dot{\alpha}}&=&-\sum_{i=1}^n\tilde{\lambda}_{\dot{\alpha}}^i
\frac{\partial}{\partial\eta^{a}_i}+\theta^{\alpha}_{a}q_{\alpha\dot{\alpha}},\nonumber\\
\mbox{4 conjugated supercharges }
\bar{Q}_{\dot{a}\dot{\alpha}}&=&-\sum_{i=1}^n\tilde{\lambda}_{\dot{\alpha}}^i
\frac{\partial}{\partial\eta^{\dot{a}}_i}+\theta^{\alpha}_{\dot{a}}q_{\alpha\dot{\alpha}}.
\end{eqnarray}
This relations imply that $\mathcal{F}_n$ takes the following form
\cite{BKV_SuperForm,HarmonyofFF}:
\begin{eqnarray}\label{F_general_structure}
\mathcal{F}_n(\{\lambda,\tilde{\lambda},\eta\},q,\theta^{a}) &=&
\delta^4(\sum_{i=1}^n \lambda_{\alpha}^i
\tilde{\lambda}_{\dot{\alpha}}^i -
q_{\alpha\dot{\alpha}})e^{\theta_{a}^{\alpha}q^{a}_{\alpha}}
\delta^4_{GR}\left(q^{\dot{a}}_{\alpha}\right)\mathcal{X}_n\left(\{\lambda,\tilde{\lambda},\eta\},q\right)\nonumber\\
\mathcal{X}_n&=&\mathcal{X}_n^{(0)} + \mathcal{X}_n^{(4)} + \ldots +
\mathcal{X}_n^{(4n-8)},
\end{eqnarray}
where
\begin{eqnarray}
q^{a}_{\alpha}=\sum_{i=1}^n\lambda_{\alpha}^i\eta^{a}_i,
~q^{\dot{a}}_{\alpha}=\sum_{i=1}^n\lambda_{\alpha}^i\eta^{\dot{a}}_i,~\delta^{4}_{GR}(q^{a/\dot{a}}_{\alpha})=\prod_{a/\dot{a},b/\dot{b}=1}^2
\epsilon^{\alpha\beta}q^{a/\dot{a}}_{\beta}q_{\alpha}^{b/\dot{b}},
\end{eqnarray}
and $\mathcal{X}^{(4m)}_n$ are the homogenous $SU(4)_R$ invariant
(more accurately $SU(2)\times SU(2)' \times U(1)$ invariant, they
can be written in $SU(4)_R$ covariant form in $\mathcal{N}=4$
harmonic superspace formulation) polynomials of the order of $4m$.

Assigning helicity $\lambda=+1$ to $|\Omega_i\rangle$ and
$\lambda=+1/2$ to $\eta$ and $\lambda=-1/2$ to
$\theta_{\alpha}^{a}$, one sees that $\mathcal{F}_n$ has an overall
helicity $\lambda_{\Sigma}=n$, $\delta^4_{GR}$ has
$\lambda_{\Sigma}=2$, exponential factor has $\lambda_{\Sigma}=0$ so
that $\mathcal{X}^{(0)}_n$ has $\lambda_{\Sigma}=n-2$,
$\mathcal{X}^{(4)}_n$ has $\lambda_{\Sigma}=n-4$, etc.
$\mathcal{X}^{(0)}_n$, $\mathcal{X}^{(4)}_n$ etc. are understood as
analogs \cite{DualConfInvForAmplitudesCorch} of the MHV, NMHV etc.
parts of superamplitude i.e. part of super form factor proportional
to $\mathcal{X}^{(0)}_n$ will contain component form factors with
overall helicity $n-2$ which we will call MHV form factors, part of
super form factor proportional to $\mathcal{X}^{(4)}_n$ will contain
component form factors with overall helicity $n-4$ which we will
call NMHV etc. up to $\mathcal{X}_n^{(4n-8)}$ overall helicity $2-n$
which we will call $\overline{\mbox{MHV}}$.

It is convenient to perform transformation from $\theta^a_{\alpha}$
to the set of axillary variables $\{\lambda^{'}_{\alpha},\eta^{'a},
\lambda^{''}_{\alpha},\eta^{''a}\}$:
\begin{equation}
\hat{T}[\ldots] = \int d^4\theta^{a}_{\alpha}
\exp(\theta_{a}^{\alpha} \sum_{i=1}^2
\lambda_{\alpha}^i\eta^{a}_i)[\ldots].
\end{equation}
After such transformation we can write $\hat{T}[\mathcal{F}_n]$ as
(let's use the notation $\lambda^{'}_{\alpha}\eta^{'a} +
\lambda^{''}_{\alpha}\eta^{''a}=\gamma^a_{\alpha}$):
\begin{equation}\label{T[superFormfactor]}
Z_n (\{\lambda,\tilde{\lambda},\eta\}, q, \{\gamma^a_{\alpha}\}) =
\hat{T}[\mathcal{F}_n] =
\delta^4(\sum_{i=1}^n\lambda_{\alpha}^i\tilde{\lambda}_{\dot{\alpha}}^i-q_{\alpha\dot{\alpha}})
\delta^4_{GR}(q^{a}_{\alpha} + \gamma^a_{\alpha}) \delta^4_{GR}
\left(q^{\dot{a}}_{\alpha}\right)\mathcal{X}_n.
\end{equation}
The algorithm of obtaining component form factors from this
supersymmetric expression was discussed in \cite{BKV_SuperForm}.

Let's make a comment about total $U(1)$ charge of $\mathcal{F}_n$
and $\hat{T}[\mathcal{F}_n]$. The $\mathcal{T}_{ab}$ operator
carries $(-4)$ charge, the $\langle\Omega_n|$ carries charge $(0)$
so the $\mathcal{F}_n$ form factor should carry $(-4)$ charge, which
indeed true and can be seen from (\ref{F_general_structure}):
$\delta^4_{GR}\left(q^{\dot{a}}_{\alpha}\right)$ carries $(-4)$
charge, while $e^{\theta_{a}^{\alpha}q^{a}_{\alpha}}$ and
$\mathcal{X}_n^{(4m)}$ are neutral. The $\hat{T}$ transformation
(the integration measure $d^4\theta^{a}_{\alpha}$) also carries
$(+4)$ charge, so that $Z_n=\hat{T}[\mathcal{F}_n]$ is neutral with
respect to $U(1)$. Note that this will be no longer true for the form factors of
operators from different supermultiplets.

The formulation of form factors discussed so far lacks of explicit
$SU(4)_R$ covariance. $SU(4)_R$ covariance can be restored in the
$\mathcal{\mathcal{N}}=4$ harmonic superspace formulation. However
such formulation does not give us any computational benefits for the
purpose of our computation and all results obtained in our none
covariant formulation can be easily translated to $SU(4)_R$
covariant formulation. We will discuss such formulation briefly in
appendix.

\subsection{Generalized unitarity for form factors at one loop}
Since we are considering the form factors of the operators 
from the stress tensor supermultiplet, such operators are protected
and do not have anomalous dimension. The reflection
of this fact for form factors at one loop will be absence of UV divergent
scalar integrals - bubbles. So in general at one 
loop\footnote{Hereafter we do not write common one
loop factor $(\lambda i\pi^{D/2}r_{\Gamma})/(2\pi)^D$ explicitly.
See appendix.} level $Z_n ^{(1)}$ can be decomposed as combination
of all possible scalar boxes $(B^{4m},
B^{3m},B^{2mh},B^{2me},B^{1m})$ and triangles 
$(T^{3m},T^{2m},T^{1m})$ integrals:
\begin{eqnarray}\label{Z_1_expantion_1}
Z_n ^{(1)}&=&\sum_i C^{4m}_iB^{4m}_i+
C^{3m}_iB^{3m}_i+C^{2mh}_iB^{2mh}_i+C^{2me}_iB^{2me}_i+C^{1m}_iB^{1m}_i\nonumber\\
&+&\sum_j C^{3m}_jT^{3m}_j+C^{2m}_jT^{2m}_j+C^{1m}_jT^{1m}_j+perm.,
\end{eqnarray}
where the sum runs over all possible distributions of the ordered
set $(p_1,\ldots,p_n)$ of individual momenta between vertexes of the
scalar integrals, while the position of the momentum $q$ carried by
the operator is fixed. $perm.$ corresponds to the cyclic
permutations of the $(p_1,\ldots,p_n)$ set of the momenta of
external particles. The latter is necessary due to the fact that
while we are considering the object that is colour ordered in the colour
space of external particles the operator is colour singlet and hence the
momentum $q$ carried by the operator can be incepted at any position
in the colour ordering \cite{FormFactorMHV_component_Brandhuber}.
This is equivalent to the consideration of all possible permutations
of external momenta, while the position of the operator momenta $q$
is fixed. In general we can write scalar box integral as:
\begin{equation}
B_{K_1^2,K_2^2,K_3^2,K_4^2}=\int
\frac{d^{D}l}{(2\pi)^{D}}\frac{1}{l^2(K_1+l)^2(K_1+K_2+l)^2(l-K_4)^2},
\end{equation}
where $\sum_{i=1}^4K^i_{\alpha\dot{\alpha}}=\sum_{i=1}^n
\lambda_{\alpha}^i \tilde{\lambda}_{\dot{\alpha}}^i -
q_{\alpha\dot{\alpha}}$. The particular scalar box integrals
$B^{4m}$, $B^{3m}$, $B^{2mh}$, $B^{2me}$, $B^{1m}$ are defined then
as: for $B^{4m}_{K_1^2,K_2^2,K_3^2,K_4^2}$ all $K_i^2\neq0$, for
$B^{3m}_{K_2^2,K_3^2,K_4^2}$ $K_1^2=0$, for $B^{2mh}_{K_3^2,K_4^2}$
$K_1^2=K_2^2=0$, for $B^{2me}_{K_2^2,K_4^2}$ $K_1^2=K_3^2=0$, for
$B^{1m}_{K_4^2}$ $K_1^2=K_2^2=K^2_3=0$. For triangle integrals we
use similar notations:
\begin{equation}
T_{K_1^2,K_2^2,K_3^2}=\int \frac{d^{D
}l}{(2\pi)^{D}}\frac{1}{l^2(K_1+l)^2(l-K_3)^2},
\end{equation}
where $\sum_{i=1}^3K^i_{\alpha\dot{\alpha}}=\sum_{i=1}^n
\lambda_{\alpha}^i \tilde{\lambda}_{\dot{\alpha}}^i -
q_{\alpha\dot{\alpha}}$, for $T^{3m}_{K_1^2,K_2^2,K_3^2}$ all
$K_i^2\neq0$, for $T^{2m}_{K_2^2,K_3^2}$ $K_1^2=0$, for
$T^{1m}_{K_3^2}$ $K_1^2=K_2^2=0$.

The dependence on the helicities of the external particles as well
as the type of operator are encoded in the $C_k$ coefficients. The
$C_k$ coefficients are Grassmann polynomials and in general, as was
explaned earlier, should have the form:
\begin{equation}
C_k \sim \delta^4_{GR}\left(q^{a}_{\alpha} +
\gamma^a_{\alpha}\right) \delta^4_{GR}
\left(q^{\dot{a}}_{\alpha}\right)\left(\mathcal{C}_k^{(0)} +
\mathcal{C}_k^{(4)} + \ldots + \mathcal{C}_k^{(4n-8)}\right),
\end{equation}
where $\mathcal{C}^{(4m)}_n$ are the homogenous $SU(4)_R$ invariant
polynomials of the order of $4m$. For example the coefficients before
scalar integrals for NMHV form factors will be proportional to
$\delta^4_{GR}\left(q^{a}_{\alpha} +
\gamma^a_{\alpha}\right)\delta^4_{GR}
\left(q^{\dot{a}}_{\alpha}\right)\mathcal{C}_k^{(4)}$. The
analytical answers for all types of one loop triangles and boxes are
known (see \cite{QCD_One_loop_Int} for review) and therefore the
problem of computation of $Z_n ^{(1)}$ reduces to the determination
of $C_k$ coefficients. The latter are computed in the unitarity
based methods by comparing the analytical properties of both sides
of the relation (\ref{Z_1_expantion_1}) viewed as the functions of
Mandelstam kinematical invariants of momenta of external particles.

To obtain the values of the coefficients $C_i$ before box integrals
it is very convenient to consider quadruple cuts
\cite{BCF,N=4onshellmethods0,Generalized_unitarity_N=4_super-amplitudes}. 
Such cuts are unique to
each box integral\footnote{Each scalar box can be uniquely specified
by its leading singularity. The latter are obtained by cutting all
four scalar propagators in the integral
\cite{BCF}.} and stop the loop momenta
flow. Therefore the quadruple cut completely determines the value of
the $C_i$ coefficient for the chosen box integral. One can
schematically write that
\begin{equation}
C_i=\frac{1}{2}\sum_{\pm S}\int \prod^4_{i}
d^2\eta_i^{a}d^2\eta_i^{\dot{a}} ~\hat{Z}^{tree}\times
\hat{\mathcal{A}}^{tree}\times \hat{\mathcal{A}}^{tree}\times
\hat{\mathcal{A}}^{tree},
\end{equation}
where $\sum_{\pm S}$ corresponds to the summation over the solutions
of the on-shell and momentum conservation conditions
\cite{Generalized_unitarity_N=4_super-amplitudes,
Triangle_coefficients}. We will not write $\sum_{\pm S}$ sums
explicitly in the most cases in the next chapters. Note that as in
the case of amplitudes
\cite{Generalized_unitarity_N=4_super-amplitudes} for the MHV and
NMHV form factors we will not need the explicit form of the
solution. Also note that the summation over the states (types of
particles) that run through the cuts is "hidden" in the Grassmann
integration ("supersums"). 
In $\mathcal{N}=4$ SYM in the case of amplitudes cuts can be evaluated 
in $D=4$ with $O(\epsilon)$ accuracy. This can be seen for example from simple
analisys based on $N=1$ superspace set up. The same arguments can be aplied also
for the form factors.
$\hat{Z}_n$ and $\hat{\mathcal{A}}_n$
correspond to form factors and amplitudes stripped from the overall
delta function $
\delta^4(\sum_{i=1}^n\lambda_{\alpha}^i\tilde{\lambda}_{\dot{\alpha}}^i-q_{\alpha\dot{\alpha}})
$ of momentum conservation. The exact types of form factors and
amplitudes entering the expression are determined by the type of
one-loop form factor (MHV, NMHV, etc.) and the particular cut we
are considering.

The situation with the coefficients before triangle scalar integrals
is a little more involved. Though triple cuts are unique to each
triangle integral the triple cut does not stop the flow of the loop
momenta completely and leaves one parameter integral $\int dt$ and 
contains contributions from different scalar box integrals.
However one can construct algorithm that allows one to extract and fix the
coefficient before \emph{particular} 
scalar triangle integral using the the triple cut
\emph{intergrand} \cite{Triangle_coefficients,WhatIsSimplestQFT}.
One can parametrise the momenta of particles
$l_i^{\alpha\dot{\alpha}}$ which crosses the cuts and associated
spinors $\lambda_{l_i},\tilde{\lambda}_{l_i}$ in terms of
combinations of external momenta and the $t$ parameter which is the
remainder of loop integral (see appendix and
\cite{Triangle_coefficients} for details). Then one can extract the coefficient
before particular triangle integral using the following relation
\begin{equation}
C_j = \mbox{Inf}_t[\frac{1}{2}\sum_{\pm S}\int \prod^3_{i}
d^2\eta_i^{a}d^2\eta_i^{\dot{a}} ~(\hat{Z}^{tree}\times
\hat{\mathcal{A}}^{tree}\times
\hat{\mathcal{A}}^{tree})](t)\Big|_{t=0},
\end{equation}
where $\mbox{Inf}_t$  means that one takes expansion in $t$ at $t
\rightarrow\infty $ and separates the term proportional to $t^0$. In
other words the coefficient before scalar triangle integral is given
by the first term in the series expansion of the corresponding
triple cut integrand in t at infinity \cite{Triangle_coefficients,
WhatIsSimplestQFT}. In the case of MHV form factors we will not need
the explicit form of the solutions $\pm S$, while the case of NMHV
form factor is more involved. We will use IR properties of the form
factors (see below) in some cases to adjust the value of the
coefficients before triangle integrals instead of the direct
computations.

\subsection{Grassmann delta functions}
Throughout this article we will need different types of the
Grassmann valued delta functions. The following notations will be
used\footnote{The relevant for us cases are $\mathcal{N}=2$,
$\mathcal{N}=4$.} for $\delta^4_{GR}$ type delta functions
introduced earlier:
\begin{equation}
\delta^8(X_{\alpha}^A)\equiv\delta^4_{GR}(X_{\alpha}^a)\delta^4_{GR}(X_{\alpha}^{\dot{a}}),
\end{equation}
while for general $\mathcal{N}$ we have:
\begin{equation}
\delta^{2\mathcal{N}}(X^{A}_{\alpha})=\prod_{A,B=1}^{\mathcal{N}}
\epsilon^{\alpha\beta}X^{A}_{\beta}X_{\alpha}^{B}.
\end{equation}
We also will use the Grassmann delta functions of another type:
\begin{equation}
\hat{\delta}^{\mathcal{N}}(\sum_i\eta^A_iC_i)=\prod_{A=1}^{\mathcal{N}}(\sum_i\eta^A_iC_i),
\end{equation}
where $A$ index runs from $1$ to $\mathcal{N}$, $\eta^A_i$ are
Grassmann variables and $C_i$ are bosonic ones. For such delta
functions we will use the following notation:
\begin{equation}
\hat{\delta}^{4}(\sum_i\eta^A_iC_i)\equiv\hat{\delta}^{2}(\sum_i\eta^a_iC_i)\hat{\delta}^{2}(\sum_i\eta^{\dot{a}}_iC_i).
\end{equation}
We will also use for saving space the notation $
d^2\eta^{a}d^2\eta^{\dot{a}}\equiv d^{4}\eta^A $ for the integration
measure. In computation of the corresponding coefficients before
scalar boxes and triangles we will be performing multiple Grassmann
integrals with Grassmann delta functions. The following relation is
extremely usefull: for some $q_{\alpha}^A$ ($A$ runs from $1$ to
$\mathcal{N}$)
$q_{\alpha}^A=\sum_{i=1}^n\lambda^{i}_{\alpha}\eta^A_i$ one can
write the following expansion over some basis
$$
q_{\alpha}^A=\lambda^{l}_{\alpha}\frac{\langle m q^A
\rangle}{\langle ml\rangle}+\lambda^{m}_{\alpha}\frac{\langle l q^A
\rangle}{\langle lm\rangle},~1\leq l\leq n,~1\leq m\leq n,~m\neq l,
$$
where $\lambda_l,\lambda_m$  should be linear independent. The
latter relation implies that
\begin{equation}\label{supersumm_delta_to_hatt_deltas}
\delta^{2\mathcal{N}}(q_{\alpha}^A)=\langle lm
\rangle^{\mathcal{N}}\hat{\delta}^{\mathcal{N}}\left(\eta^A_{l}+\sum_{i=1}^n\frac{\langle
mi \rangle}{\langle ml
\rangle}\eta_i^A\right)\hat{\delta}^{\mathcal{N}}\left(\eta^A_{m}+\sum_{i=1}^n\frac{\langle
li \rangle}{\langle lm \rangle}\eta_i^A\right), ~i \neq l,~i\neq m.
\end{equation}
For example using this relation one can immediately show that
\cite{BKV_SuperForm}
\begin{eqnarray}\label{supersummGeneralTHeory}
&&\int d^\mathcal{N} \eta_{l_1} d^\mathcal{N}\eta_{l_2}
\delta^{2\mathcal{N}}
\left(\lambda_{\alpha}^{l_1}\eta_{l_1}^A+\lambda_{\alpha}^{l_2}\eta_{l_2}^A+Q_{\alpha}^A\right)
\delta^{2\mathcal{N}} \left(\lambda_{\alpha}^{l_1}\eta_{l_1}^A+\lambda_{\alpha}^{l_2}\eta_{l_2}^A-P_{\alpha}^A\right)\nonumber\\
&&=\langle l_1l_2\rangle^{\mathcal{N}}
\delta^{2\mathcal{N}}\left(P_{\alpha}^A+Q_{\alpha}^A\right),
\end{eqnarray}
which is important relation for the two particle supersums.

We also want to note that the the computation of integrals over
$\int d^2\eta_i^{a}$ and $\int d^2\eta_i^{\dot{a}}$ in the quadruple
and triple cuts may be different in details, but can be performed in
such a way that leads to the same bosonic coefficient and slightly
different Grassmann delta functions. One can formulate the following
rule to simplify computations: one takes integrals over $\int
d^2\eta_i^{a}d^2\eta_i^{\dot{a}}$ formally replacing
\begin{equation}\label{integration_trick}
\delta^4_{GR}(X_{\alpha}^a+\gamma^a_{\alpha})\delta^4_{GR}(X_{\alpha}^{\dot{a}})
\rightarrow \delta^8(X_{\alpha}^A+\gamma^A_{\alpha})
\end{equation}
and integrating over $\int d^4\eta_i^{A}$. After the integration we
have to put $\gamma^{\dot{a}}_{\alpha}=0$.

\subsection{Tree level $\mbox{MHV}$ and $\overline{\mbox{MHV}}$ amplitudes and form factors}
We will need as the building blocks in the computation of the
coefficients before scalar integrals in the MHV and NMHV case at one
loop level several explicit expressions for the tree level form
factors \cite{BKV_SuperForm,HarmonyofFF} and amplitudes
\cite{MHV_first,Generalized_unitarity_N=4_super-amplitudes}:
\begin{equation}
Z^{tree,MHV}_n=
\delta^4(\sum_{i=1}^n\lambda^{\alpha}_i\tilde{\lambda}^{\dot{\alpha}}_i+q^{\alpha\dot{\alpha}})
\frac{\delta^4_{GR}\left(q^{a}_{\alpha} + \gamma^a_{\alpha}\right)
\delta^4_{GR} \left(q^{\dot{a}}_{\alpha}\right)}{\langle12\rangle
\ldots \langle n1\rangle}.
\end{equation}
This expression is valid for any $n \geq 2$ without any kinematical
constraints on $\lambda_i,\tilde{\lambda}_i$ variables. For the MHV
amplitudes we have:
\begin{equation}
\mathcal{A}_n^{tree,~MHV}=\delta^4(\sum_{i=1}^n\lambda^{\alpha}_i\tilde{\lambda}^{\dot{\alpha}}_i)
\frac{\delta^8(q^A_{\alpha})}{\langle12\rangle \ldots \langle
n1\rangle}.
\end{equation}
This expression is valid for any $n \geq 4$ without any kinematical
constraints. While for $n=3$ this expression exists only for the
complex values of the momenta\footnote{One can make analytical
continuation to the real values of the  momenta of external
particles in the final expressions.} in $(+---)$ signature. This
implies the following kinematical constraints on
$\lambda_i,\tilde{\lambda}_i$ variables
\cite{Generalized_unitarity_N=4_super-amplitudes}:
\begin{equation}
\mbox{MHV}_3:~\tilde{\lambda}_i\langle
ik\rangle=-\tilde{\lambda}_j\langle
jk\rangle,~[ij]=0,~\mbox{any}~i,j,k ~\mbox{at the same vertex}.
\end{equation}
We will also need the expression for the three point
$\overline{\mbox{MHV}}_3$ amplitude 
\cite{N=4onshellmethods0,Generalized_unitarity_N=4_super-amplitudes}:
\begin{equation}
\mathcal{A}_3^{tree,~\overline{MHV}}=\delta^4(\sum_{i=1}^3\lambda^{\alpha}_i\tilde{\lambda}^{\dot{\alpha}}_i)
\frac{\hat{\delta}^4(\eta_1^A[23]+\mbox{cycl.perm.})}{[12] [23]
[31]}.
\end{equation}
This expression also exists only for the complex values of the
momenta, that implies the following constraints:
$$
\overline{\mbox{MHV}}_3:~\lambda_i[ ik]=-\lambda_j[ jk],~\langle
ij\rangle=0,~\mbox{any}~i,j,k ~\mbox{at the same vertex}.
$$
Note that this expression is $4$'th degree in $\eta$'s while all MHV
amplitudes and form factors are of the $8$'th degree. We also will
need the form of $Z_3^{tree,~\overline{MHV}}$ which is NMHV form
factor at the same time in full analogy with $n=5$ point amplitude.
We will discuss the structure of it in separate section later on.

\begin{figure}[t]
 \begin{center}
  \epsfxsize=12cm
 \epsffile{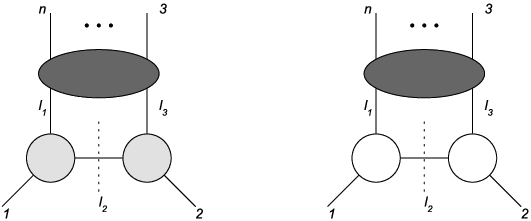}
 \end{center}\vspace{-0.2cm}
 \caption{Vanishing configurations of $\mbox{MHV}_3$ (grey) and
 $\overline{\mbox{MHV}}_3$ (white)
vertexes. Dark grey blob corresponds to other parts of
"diagram".}\label{fig_zero_cofig}
 \end{figure}

The discussed above kinematical constrains on
$\lambda_i,\tilde{\lambda}_i$ spinors immediately lead to the fact
that some configurations of MHV and $\overline{\mbox{MHV}}$ vertexes
give vanishing result
\cite{Generalized_unitarity_N=4_super-amplitudes}. See
fig.\ref{fig_zero_cofig}.

\section{MHV warm-up}
\begin{figure}[t]
 \begin{center}
  \epsfxsize=13cm
 \epsffile{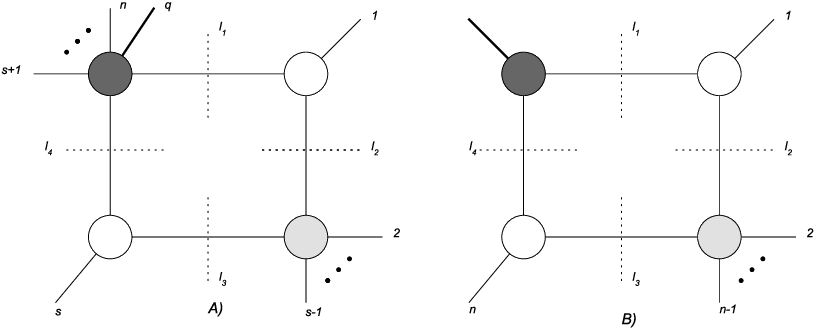}
 \end{center}\vspace{-0.2cm}
 \caption{All possible cuts for the box scalar integrals to MHV form factor. Dark grey vertex is MHV form factor,
 grey vertex is MHV amplitude, white vertex is $\overline{\mbox{MHV}}_3$ amplitude.}\label{MHV_quadruple_cuts}
 \end{figure}

\begin{figure}[t]
 \begin{center}
  \epsfxsize=12cm
 \epsffile{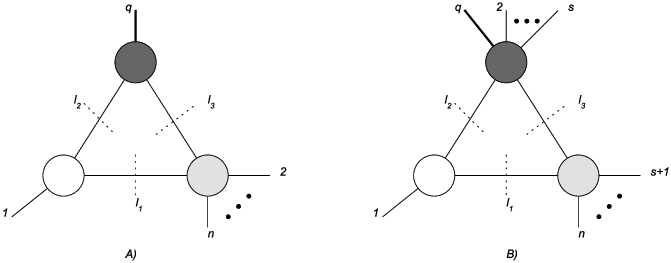}
 \end{center}\vspace{-0.2cm}
 \caption{All possible cuts for the triangle scalar integrals to MHV form factor.}\label{MHV_triple_cuts}
 \end{figure}

As a warm-up before computations of NMHV form factors we will
discuss how generalized unitarity works in MHV case. MHV form
factors should be the lowest components in $\eta$'s expansion of
$Z_n$ of Grassmann degree 8. This implies the following
configurations of MHV and $\overline{\mbox{MHV}}$ vertexes
$$
\mbox{MHV}\times\mbox{MHV}\times\overline{\mbox{MHV}}_3\times\overline{\mbox{MHV}}_3
$$
for the quadruple cut integrand, and
$$
\mbox{MHV}\times\mbox{MHV}\times\overline{\mbox{MHV}}_3
$$
for the triple cut integrand. Taking into account that
configurations of vertexes depicted on fig.\ref{fig_zero_cofig}
vanish we conclude that the only contributing cuts are those
depicted on fig.\ref{MHV_quadruple_cuts} and
fig.\ref{MHV_triple_cuts}.
They correspond to the $B^{2me}$ (and $B^{1m}$ as the limiting case)
and $T^{2m}$ scalar integrals. So the MHV form factor should have
the form
\begin{equation}\label{Z_1_expantion}
Z_n ^{(1),MHV}=\sum_i C_i^{2me} B^{2me}_i+\sum_j C_j^{2m}
~T^{2m}_j+perm.,
\end{equation}
($B^{1m}$ scalar integrals can also appear as the limiting case of
$B^{2me}$). Let's start with the quadruple cuts for $B^{2me}$
integral. I.e. we are going to compute the corresponding $C^{2me}$
coefficient before such integral. This computation is very similar
to those in \cite{Generalized_unitarity_N=4_super-amplitudes}. We
will treat the configurations where the vertex
$\hat{Z}_2^{tree,MHV}$ is present and configurations with
$\hat{Z}_n^{tree,MHV}$, $n \geq 3$ separately. The usefulness of
such treatment will be clear later on. We have for the configuration
$A)$ which corresponds to $B^{2me}_{s_{2...s-1},s_{1...s}}$ scalar
integral\footnote{We will suppress $\alpha$ $SL(2,\mathbb{C})$
indexes indices in arguments of Grassmann delta functions in some
cases.} (the notation
\begin{equation}
s_{r...s-1}=\left(\sum_{i=r}^{s-1}p_i\right)^2=p_{r...s-1}^2
\end{equation}
is used):
\begin{eqnarray}
&C^{2me}_{s_{2...s-1},s_{1...s}}&=\int\prod_{i=1}^4d^4\eta^A_{l_{i}}
\frac{\hat{\delta}^4(\eta_1[l_2l_1]-\eta_{l_2}[l_11]+\eta_{l_1}[1l_2])}{[1l_2][l_2l_1][l_11]}
\frac{\hat{\delta}^4(\eta_s[l_4l_3]-\eta_{l_4}[l_3s]+\eta_{l_3}[sl_4])}{[sl_4][l_4l_3][l_3s]}\nonumber\\
&\times&\frac{\delta^4_{GR}(\sum_{i=s+1}^n\lambda_i\eta_i^{a}+\gamma^{a}+\lambda_{l_4}\eta_{l_4}^a-\lambda_{l_1}\eta_{l_1}^a)
\delta^4_{GR}(\sum_{i=s+1}^n\lambda_i\eta_i^{\dot{a}}+\lambda_{l_4}\eta_{l_4}^{\dot{a}}-\lambda_{l_1}\eta_{l_1}^{\dot{a}})}{\langle
l_1l_4\rangle \ldots \langle nl_1\rangle}\nonumber\\
&\times&\frac{\delta^8(\sum_{i=2}^{s-1}\lambda_i\eta^A_i+\lambda_{l_2}\eta_{l_2}^A-\lambda_{l_3}\eta_{l_3}^A)}{\langle
l_2 l_3 \rangle \ldots \langle 2 l_2 \rangle}.\nonumber\\
\end{eqnarray}
Performing Grassmann integration with first two $\hat{\delta}^4$ and
then integrating the remaining delta functions with the use of
(\ref{supersumm_delta_to_hatt_deltas}) and kinematical constraints
of the $\overline{\mbox{MHV}}$ vertexes ((\ref{integration_trick})
also can be used to simplify algebraic manipulations) we obtain
\begin{eqnarray}
C_{s_{2...s-1},s_{1...s}}^{2me}&=&Z^{tree,MHV}_n\frac{\langle12\rangle\langle
s-1s\rangle\langle ss+1\rangle\langle
n1\rangle[1|l_1l_3|s]^4}{[1|l_1l_4|s]~[1|l_2l_3|s]~[l_2l_1][l_4l_3]~\langle
s-1l_3\rangle\langle2l_2\rangle~[l_3l_4]~\langle
nl_1\rangle\langle s+1l_4\rangle}.\nonumber\\
\end{eqnarray}
This expression can be transformed with the use of momentum
conservation conditions in each vertex and $\overline{\mbox{MHV}}$
kinematical constraints to the form
\begin{eqnarray}
C^{2me}_{s_{2...s-1},s_{1...s}}&=&Z^{tree,MHV}_n\frac{1}{2}Tr(l_2l_1l_4l_3)=\nonumber\\
&=&Z^{tree,MHV}_n\frac{1}{2}((l_1-l_4)^2(l_3-l_2)^2-(l_1-l_3)^2(l_4-l_2)^2).
\end{eqnarray}
This result is identical to the MHV amplitude case
\cite{Generalized_unitarity_N=4_super-amplitudes}. Note also that as
was in the amplitude case the $\sum_{\pm S}$ can be evaluated
without the use of explicit solutions for $l_i$. So we have that the
coefficient before $B^{2me}_{s_{2...s-1},s_{1...s}}$ scalar integral
takes the form:
\begin{eqnarray}\label{coefficients_Box_MHV_1}
C_{s_{2...s-1},s_{1...s}}^{2me}&=&Z^{tree,MHV}_n\frac{1}{2}\Delta^{2me}_{s_{2...s-1},s_{1...s}}\nonumber\\
\Delta^{2me}_{s_{2...s-1},s_{1...s}}&=&s_{2...s-1}s_{1...s}-s_{1...s-1}s_{2...s}.
\end{eqnarray}
The case of configuration $B)$ which corresponds to the
$B^{2me}_{s_{2...n-1},q^2}$ scalar integral is slightly different in
computational details:
\begin{eqnarray}
&C_{s_{2...n-1},q^2}^{2me}&=\int\prod_{i=1}^4d^4\eta^A_{l_{i}}
\frac{\hat{\delta}^4(\eta_1[l_2l_1]-\eta_{l_2}[l_11]+\eta_{l_1}[1l_2])}{[1l_2][l_2l_1][l_11]}
\frac{\hat{\delta}^4(\eta_n[l_4l_3]-\eta_{l_4}[l_3n]+\eta_{l_3}[nl_4])}{[nl_4][l_4l_3][l_3n]}\nonumber\\
&\times&\frac{\delta^4_{GR}(\sum_{-l_1,l_4}\lambda_i\eta_i^{a}+\gamma^{a})\delta^4_{GR}(\sum_{-l_1,l_4}\lambda_i\eta_i^{\dot{a}})}{\langle
l_1l_4\rangle^2}
\frac{\delta^8(\sum_{i=2}^{n-1}\lambda_i\eta^A_i+\lambda_{l_2}\eta_{l_2}^A-\lambda_{l_3}\eta_{l_3}^A)}{\langle
l_2 l_3 \rangle \ldots \langle 2 l_2 \rangle},\nonumber\\
\end{eqnarray}
but leads to essentially the same final result:
\begin{eqnarray}
C_{s_{2...n-1},q^2}^{2me}&=&Z^{tree,MHV}_n\frac{\langle12\rangle\langle
n-1n\rangle\langle
n1\rangle[1|l_1l_4|n]^4}{[1|l_1l_4|n]~[1l_2][l_2l_1]~\langle
n-1l_3\rangle\langle
l_3l_2 \rangle\langle2l_2\rangle~[l_3l_4][l_3n]~\langle l_1l_4\rangle},\nonumber\\
\end{eqnarray}
which, as in the previous case, can be simplified so that the
coefficient before $B^{2me}_{s_{2...n-1},q^2}$ scalar integral takes
the form:
\begin{eqnarray}\label{coefficients_Box_MHV_2}
C_{s_{2...n-1},q^2}^{2me}&=&Z^{tree,MHV}_n\frac{1}{2}\Delta^{2me}_{s_{2...n-1},q^2}\nonumber\\
\Delta^{2me}_{s_{2...n-1},q^2}&=&s_{2...n-1}q^2-s_{1...n-1}s_{2...n}.
\end{eqnarray}

Now let's consider triple cuts. Let's consider the $A)$
configuration which corresponds to the coefficient before
$T^{2m}_{s_{2...n},q^2}$ scalar integral. We have for the
corresponding triple cut integrand
\begin{eqnarray}
&&\int\prod_{i=1}^3d^4\eta^A_{l_{i}}\frac{\hat{\delta}^4(\eta_1[l_2l_1]-\eta_{l_2}[l_11]+\eta_{l_1}[1l_2])}{[1l_2][l_2l_1][l_11]}
\frac{\delta^4_{GR}(\sum_{l_2,l_3}\lambda_i\eta_i^{a}+\gamma^{a})\delta^4_{GR}(\sum_{l_2,l_3}\lambda_i\eta_i^{\dot{a}})}{\langle
l_2l_3\rangle^2}\nonumber\\&\times&\frac{\delta^8(\sum_{i=2}^{n}\lambda_i\eta^A_i+\lambda_{l_3}\eta_{l_3}^A-\lambda_{l_1}\eta_{l_1}^A)}{\langle
l_1 l_3 \rangle \ldots \langle n l_1 \rangle}.
\end{eqnarray}
Performing Grassmann integration in the same fashion as in the
quadruple cut cases (first integrating $\hat{\delta}^2$, then with
the use of (\ref{supersumm_delta_to_hatt_deltas}) and kinematical
constraints of the $\overline{\mbox{MHV}}$ vertex integrating
$\delta^8$) we obtain for the triple cut integrand:
\begin{eqnarray}
Z_n^{tree,MHV}\frac{\langle 12 \rangle \langle
n1\rangle[1|l_1|l_3\rangle^4}{\langle l_3l_2
\rangle^2~[1l_2][l_2l_1][l_11]~\langle nl_1\rangle\langle l_1l_3
\rangle \langle l_32 \rangle }.
\end{eqnarray}
This expression can be further simplified with the use of the
momentum conservation conditions associated with each vertex to the
form\footnote{We use the following notations for the scalar products
of momenta of external particles $(p_{i\ldots l},p_{j\ldots
k})=(i+\ldots+l,j+\ldots+k)$.}:
\begin{eqnarray}
Z_n^{tree,MHV}\frac{Tr(2l_2l_1l_3)}{(l_32)},
\end{eqnarray}
where the trace can be evaluated, with the use of momentum
conservation conditions and kinematical constraints of the
$\overline{\mbox{MHV}}$ vertex to the form
\begin{eqnarray}
Z_n^{tree,MHV}\frac{1}{4}\left(2(s_{2...n}-q^2)+\sum_{\pm
S}\frac{(q2)s_{2...n}-(2,1+q)q^2}{(l_32)}\right).
\end{eqnarray}
Now applying the following parametrization for
$l_i^{\alpha\dot{\alpha}}$:
$l_i^{\alpha\dot{\alpha}}=A_1^{\alpha\dot{\alpha}}+tA_2^{\alpha\dot{\alpha}}+1/tA_3^{\alpha\dot{\alpha}}$,
where $A_{1,2,3}^{\alpha\dot{\alpha}}$ are some constants that
depend on external momenta (see \cite{Triangle_coefficients} and
appendix for details), we see that
\begin{eqnarray}\label{coefficients_Tri_MHV}
C_{s_{2...n},q^2}^{2m} &=&
Z_n^{tree,MHV}\frac{1}{4}\mbox{Inf}_t[2(s_{2...n}-q^2)+\sum_{\pm
S}\frac{(q2)s_{2...n}-(2,1+q)q^2}{(l_32)}]\Big|_{t=0}\nonumber\\
&=&Z_n^{tree,MHV}\frac{1}{2}(s_{2...n}-q^2),
\end{eqnarray}
so that the coefficient before $T_{s_{2...n},q^2}^{2m}$ triangle
scalar integral takes the form:
\begin{eqnarray}
C_{s_{2...n},q^2}^{2m}&=&Z^{tree,MHV}_n\frac{1}{2}\Delta_{s_{2...n},q^2}^{2m}\nonumber\\
\Delta_{s_{2...n},q^2}^{2m}&=&s_{2...n}-q^2.
\end{eqnarray}
Now let's consider the $B)$ configuration which corresponds to the
coefficient before scalar integral $T^{2m}_{s_{s+1...n},s_{1...n}}$.
We have for the corresponding triple cut integrand
\begin{eqnarray}
&&\int\prod_{i=1}^3d^4\eta^A_{l_{i}}\frac{\hat{\delta}^4(\eta_1[l_2l_1]-\eta_{l_2}[l_11]+\eta_{l_1}[1l_2])}{[1l_2][l_2l_1][l_11]}
\frac{\delta^8(\sum_{i=s+1}^{n}\lambda_i\eta^A_i+\lambda_{l_3}\eta_{l_3}^A-\lambda_{l_1}\eta_{l_1}^A)}{\langle
l_1 l_3 \rangle \ldots \langle n l_1
\rangle}\nonumber\\&\times&\frac{\delta^4_{GR}(\sum_{i=2}^s\lambda_i\eta_i^{a}+\gamma^{a}+\lambda_{l_2}\eta^{a}_{l_2}-\lambda_{l_3}\eta^{a}_{l_3})\delta^4_{GR}(\sum_{i=2}^s\lambda_i\eta_i^{\dot{a}}+\lambda_{l_2}\eta^{\dot{a}}_{l_2}-\lambda_{l_3}\eta^{\dot{a}}_{l_3})}{\langle
l_2l_3\rangle^2}.
\end{eqnarray}
Performing the Grassmann integration with delta functions we obtain:
\begin{eqnarray}
Z_n^{tree,MHV}\frac{\langle 12 \rangle \langle n1\rangle \langle
ss+1 \rangle[1|l_1|l_3\rangle^4}{\langle l_22 \rangle \langle
sl_3\rangle \langle l_3l_2\rangle~[1l_2][l_2l_1][l_11]~\langle
l_3s+1\rangle\langle nl_1 \rangle \langle l_1l_3 \rangle }.
\end{eqnarray}
This expression can be simplified with the use of the momentum
conservation conditions associated with each vertex to the form
\begin{eqnarray}
Z_n^{tree,MHV}\frac{1}{4}\sum_{\pm
S}\left(\frac{Tr(l_1l_2s+1l_3)}{(l_3s+1)}-\frac{Tr(l_1l_2sl_3)}{(l_3s)}\right).
\end{eqnarray}
The traces can be evaluated as in the previous case and we get
\begin{eqnarray}
&&Z_n^{tree,MHV}\frac{1}{4}\sum_{\pm
S}\left(\frac{D_{1s+1}}{(l_3s+1)}+\frac{D_{1s}}{(l_3s)}\right),\nonumber\\
&&D_{1j}=(1,1+\sum_{k=s+1}^{n}k)(j,1+\sum_{k=s+1}^{n}k)-(1j)(1+\sum_{k=s+1}^{n}k)^2.\nonumber\\
\end{eqnarray}
Note that $D_{1j}$ depends only on the external momenta. So using
the parametrization for $l_i^{\alpha\dot{\alpha}}$:
$l_i^{\alpha\dot{\alpha}}=A_1^{\alpha\dot{\alpha}}+tA_2^{\alpha\dot{\alpha}}+1/tA_3^{\alpha\dot{\alpha}}$,
where $A_{1,2,3}^{\alpha\dot{\alpha}}$ are some constants that
depend on external momenta, we see that
\begin{equation}
C_{s_{s+1...n},s_{1...n}}^{2m} = Z_n^{tree,MHV}\frac{1}{4}\sum_{\pm
S}\mbox{Inf}_t[\frac{D_{1s+1}}{(l_3s+1)}+\frac{D_{1s}}{(l_3s)}]\Big|_{t=0}=0.
\end{equation}
So we conclude that potentially contributing integrals
$T^{2m}_{s_{s+1n},s_{1n}}$ does not actually appear. This fact was
first established in \cite{FormFactorMHV_component_Brandhuber}.

The latter fact may be puzzling. The resolution of this puzzle comes
from the observation that $\hat{Z}_2^{tree,MHV}$ and
$\hat{Z}_n^{tree,MHV},~n>2$ vertexes that enter $A)$ and $B)$ triple
cuts have different number of $\lambda_{l_i}$ spinors in
denominators, so that the $\hat{Z}_2^{tree,MHV}$ vertex is in some
sense singled out (that's why we treated $A)$ and $B)$ cases
separately for the quadruple cuts)
$$
Z_2^{tree,MHV}~\sim\frac{1}{\langle
l_2l_3\rangle^2},~Z_n^{tree,MHV}~\sim\frac{1}{\langle il_2
\rangle\langle l_2l_3\rangle \langle j\l_3\rangle}.
$$
Using explicit solutions for $\lambda_{l_i}$ in terms of t and
external momenta \cite{Triangle_coefficients} for $T^{2m}$ triangles
one sees that
$$
\frac{1}{\langle l_2l_3\rangle^2}\sim t^0,
$$
so if we take into account the whole expression we have none
vanishing contribution in $t \rightarrow \infty$ limit, while
$$
\frac{1}{\langle il_2 \rangle\langle l_2l_3\rangle \langle
j\l_3\rangle}\sim t^{-1},
$$
so in the $t \rightarrow \infty$ limit we get zero. We verified that
this
is the general pattern for MHV and NMHV cases for the $T^{2m}$ and $T^{3m}$ integrals,
so on general grounds \emph{we can conclude that the only allowed
triangle integrals should necessary contain one massive} $q^2$
\emph{leg} in MHV and NMHV sectors. This fact reduces the number of
necessary triple cuts significantly. 
Note, that the properties
of $Z_2^{tree,MHV}$ vertex in the $t \rightarrow \infty$ limit
resemble those of the $z \rightarrow \infty$ limit in the BCFW
recursion. 
%
%
%

We see now that the contributing type of scalar integrals in the MHV
case are $B^{2me}_{s_{2...s-1},s_{1...s}},~T^{2m}_{s_{2...n},q^2}$
(and $B^{2me}_{s_{2...n-1},q^2}$ as the limiting case of
$B^{2me}_{s_{2...s-1},s_{1...s}}$). The coefficients before these
integrals are given correspondingly by
(\ref{coefficients_Box_MHV_1}) and (\ref{coefficients_Tri_MHV}).

Note also that the combinations of scalar products of external
momenta $\Delta$'s in the coefficients before corresponding scalar
integrals (boxes and triangles) match the scalar integral in such
way that the coefficients before the $1/\epsilon^2$ IR pole (the use
of the dimensional regularization is implied) will have the form
$\sim\Delta^{-1}/\epsilon^2$ (see appendix). This allows us to
define dimensionless functions $\mathcal{B}^{i}$ and
$\mathcal{T}^{i}$, just as in the case of amplitudes
\cite{Generalized_unitarity_N=4_super-amplitudes},  for all types of
boxes and triangles as the result of evaluation of the corresponding
scalar integral through $O(\epsilon)$ multiplied by the
corresponding $\Delta$ coefficient.

As an example, for three point MHV form factor $Z_3^{(1),MHV}$ one
can obtain, in precise agreement with
\cite{FormFactorMHV_component_Brandhuber,BKV_SuperForm}:
\begin{eqnarray}
Z_{3}^{(1),MHV}/Z_{3}^{tree,MHV}&=&\frac{1}{2}\mathcal{B}^{1m}(1,2,3|q^2)+\frac{1}{2}\mathcal{B}^{1m}(1,3,2|q^2)+\frac{1}{2}\mathcal{B}^{1m}(2,1,3|q^2)
\nonumber\\
&+&\mathcal{T}^{2m}(1|q^2,(2+3)^2)+
\mathcal{T}^{2m}(2|q^2,(1+3)^2) \nonumber\\
&+&\mathcal{T}^{2m}(3|q^2,(1+2)^2).
\end{eqnarray}
Here we write the ordering of massive/massless legs explicitly,
using the convention:
$$
\mathcal{G}^{i}(\mbox{massless legs}|(\mbox{massive legs})^2),
$$
where $\mathcal{G}^{i}$ is dimensionless function based on the type
of scalar integral under consideration. The IR divergent part of
this result is given by:
\begin{eqnarray}
Z_{3}^{(1),MHV}/Z_{3}^{tree,MHV}\Big|_{IR}&=&\frac{1}{\epsilon^2}\sum_{i=1}^3\left(\frac{s_{ii+1}}{\mu^2}\right)^{\epsilon},
\end{eqnarray}
while finite part is:
\begin{eqnarray}
Z_{3}^{(1),MHV}/Z_{3}^{tree,MHV}\Big|_{fin}&=&-\mbox{Li}_2\left(1-\frac{q^2}{s_{12}}\right)
-\mbox{Li}_2\left(1-\frac{q^2}{s_{23}}\right)
-\mbox{Li}_2\left(1-\frac{q^2}{s_{31}}\right)\nonumber\\
&-&\frac{1}{2}\mbox{Log}\left(\frac{s_{12}}{s_{23}}\right)-
\frac{1}{2}\mbox{Log}\left(\frac{s_{23}}{s_{31}}\right)-
\frac{1}{2}\mbox{Log}\left(\frac{s_{31}}{s_{12}}\right)-\frac{\pi^2}{2}.\nonumber\\
\end{eqnarray}

Note that the IR divergences in the sum of
$B^{2me}_{s_{2...s-1},s_{1...s}},~T^{2m}_{s_{2...n},q^2}$ will
always \cite{FormFactorMHV_component_Brandhuber}  combine in such a
way that
\begin{equation}
Z_n^{(1),MHV}\Big|_{IR}=Z_n^{tree,MHV}\frac{1}{\epsilon^2}\sum_{i=1}^n\left(\frac{s_{ii+1}}{\mu^2}\right)^{\epsilon}.
\end{equation}
This is in fact the consequence of the fact that IR poles should
cancel in IR finite observables \cite{KLN} such as inclusive cross
sections \cite{EKSglu, EKS, KunsztSoper, finite1,finite2}, energy
flow functions \cite{EF, EFK, hofmal}, etc. based on form factors
\cite{vanNeerven:1985ja, Sasha} and we expect that similar behavior
will take place for all types (MHV, NMHV, etc) of form factors
\begin{equation}\label{IR_consist}
Z_n^{(1)}\Big|_{IR}=Z_n^{tree}\frac{1}{\epsilon^2}\sum_{i=1}^n\left(\frac{s_{ii+1}}{\mu^2}\right)^{\epsilon}.
\end{equation}

It was noticed in \cite{FormFactorMHV_component_Brandhuber} that it
is likely possible to find the perturbative description of the form
factor in terms of the periodic Wilson loops. This fact suggests
that the use of the dual variables
\cite{DualConfInvForAmplitudesCorch} as in the amplitude case will
be useful though the dual conformal properties
\cite{BKV_FormFN=1,BKV_SuperForm} of form factors remain obscure. We
will use dual variables as compact notations in some cases.
\begin{figure}[t]
 \begin{center}
  \epsfxsize=15cm
 \epsffile{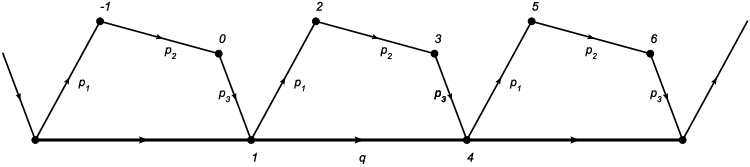}
 \end{center}\vspace{-0.2cm}
 \caption{Example of the dual contur for the MHV n=3 form factor.}\label{fig_dual_contur}
 \end{figure}
One introduces dual coordinates $x^{\alpha\dot{\alpha}}_i$ as
\begin{equation}
x_{rs}^{\alpha\dot{\alpha}}=\sum_{i=r}^{s-1}p_i^{\alpha\dot{\alpha}}.
\end{equation}
Note that in the case of the periodic contour within one period
there are $n$ labels for the momenta of external particles but $n+1$
label for the dual $x_i$ points. For example we can write the
momentum carried by the operator $q$ as
$q=\sum_{i=1}^np_i=x_{1n+1}$. See fig.\ref{fig_dual_contur} for the
$n=3$ case. Note also that for some kinematical invariants we will
need labels from different periods. For example
$s_{31}=(1+3)^2=x_{35}^2$. Moreover one kinematical invariant can
have different representations from $x$'s from different periods.
For example the following identity holds:
$s_{31}=(1+3)^2=x_{35}^2=x^2_{02}$.

In such notations we have for the coefficients before
$B^{2me}_{s_{2...s-1},s_{1...s}},~T^{2m}_{s_{2...n},q^2}$ scalar
integrals:
\begin{eqnarray}\label{coefficients_Box_MHV_1_dual}
C_{s_{2...s-1},s_{1...s}}^{2me}&=&Z^{tree,MHV}_n\frac{1}{2}\Delta^{2me}_{s_{2...s-1},s_{1...s}}\nonumber\\
\Delta^{2me}_{s_{2...s-1},s_{1...s}}&=&x^2_{2s}x_{1s+1}^2-x^2_{1s}x_{2s+1}^2,
\end{eqnarray}
\begin{eqnarray}
C_{s_{2...n},q^2}^{2m}&=&Z^{tree,MHV}_n\frac{1}{2}\Delta_{s_{2...n},q^2}^{2m}\nonumber\\
\Delta_{s_{2...n},q^2}^{2m}&=&x^2_{2n+1}-x_{1n+1}^2.
\end{eqnarray}
For example, the finite part of three point MHV form factor can be
written in terms of dual variables as:
\begin{eqnarray}\label{MHV_3_fin_part_Dual_var}
Z_{3}^{(1),MHV}/Z_{3}^{tree,MHV}\Big|_{fin}=-(1+\mathbb{P}+\mathbb{P}^2)
\left(\mbox{Li}_2\left(1-\frac{x_{14}^2}{x^2_{13}}\right)+
\frac{1}{2}\mbox{Log}^2\left(\frac{x^2_{13}}{x^2_{24}}\right)+\frac{\pi^2}{6}\right),
\nonumber\\
\end{eqnarray}
here $\mathbb{P}$ is permutation operator which acts on the momenta
or dual variables labels. Note that there are no periodicity
conditions on indices of $x$ dual coordinates. We will use such
variables for the coefficients before scalar integrals also in the
NMHV case which we are going to discuss now.
\section{NMHV form factor}
\begin{figure}[t]
 \begin{center}
  \epsfxsize=7cm
 \epsffile{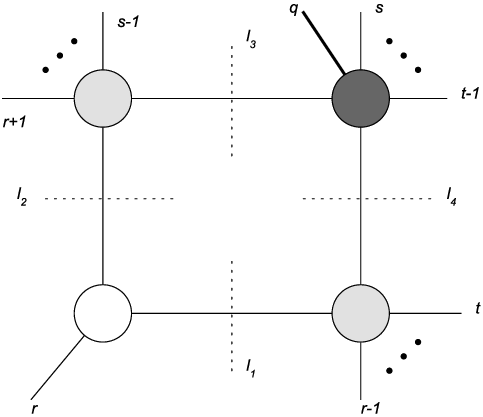}
 \end{center}\vspace{-0.2cm}
 \caption{Diagrammatical representation of the $R^{(1)}_{rst}$. }\label{NMHV_R1}
 \end{figure}
\begin{figure}[t]
 \begin{center}
  \epsfxsize=7cm
 \epsffile{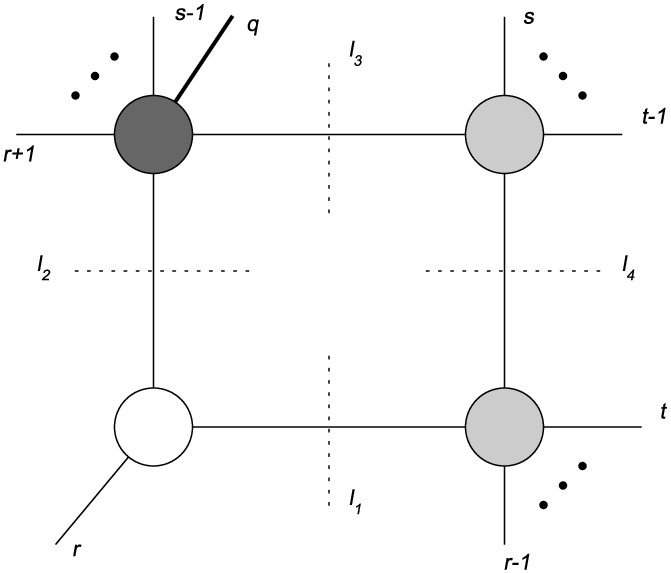}
 \end{center}\vspace{-0.2cm}
 \caption{Diagrammatical representation of the $R^{(2)}_{rst}$. }\label{NMHV_R2}
 \end{figure}
For NMHV form factors in general we have the folloving expantion in
terms of scalar integrals:
\begin{equation}\label{Z_NMHV_expantion}
Z_n ^{(1),NMHV}=\sum_i
C_i^{2me}B^{2me}_i+C_i^{2mh}B^{2mh}_i+C_i^{3m}B^{3m}_i +\sum_j
C_j^{2m}T^{2m}_j+C_j^{3m}T^{3m}_j+perm.,
\end{equation}
($B^{1m}$ scalar integrals can also appear as the limiting case of
$B^{2me}$). NMHV form factors should be the next-to the lowest
components in $\eta$'s expansion of $Z_n$ of Grassmann degree 8+4,
so the Grassmann degree of the $C$ coefficients are 12. This implies
the following configurations of MHV and $\overline{\mbox{MHV}}$
vertexes
$$
\mbox{MHV}\times\mbox{MHV}\times\mbox{MHV}\times\overline{\mbox{MHV}}_3,
$$
and
$$
\mbox{NMHV}\times\mbox{MHV}\times\overline{\mbox{MHV}}_3\times\overline{\mbox{MHV}}_3,
$$
for the quadruple cut integrands which defines the coefficients
before box type scalar integrals. The configuration involving NMHV
vertex (which can be amplitude or form factor) can be treated
recursively. To compute n point NMHV form factor at one loop one
will need $n-1$ point NMHV tree form factor (all the tree
$\mbox{NMHV}$ amplitudes are known at least in principle in the form
which can be used in our computations
\cite{DualConfInvForAmplitudesCorch,Generalized_unitarity_N=4_super-amplitudes}).
One can extract $n-1$ point NMHV tree form factor from $n-1$ point
NMHV one loop form factor using (\ref{IR_consist}). The recursion
starts with the $\overline{\mbox{MHV}}_3$ form factor which is also
$\mbox{NMHV}_3$ one. In the $\mbox{NMHV}_{3}$ case there are no
contributions from
$\mbox{NMHV}\times\mbox{MHV}\times\overline{\mbox{MHV}}_3\times\overline{\mbox{MHV}}_3$
configuration. So one can extract from (\ref{IR_consist}) the form
of $\mbox{NMHV}_{3}$ at tree level and then use it in the
computations of $\mbox{NMHV}_{4}$ from which using
(\ref{IR_consist}) one can extract $\mbox{NMHV}_{4}$ at tree level
and then use it in the computations of $\mbox{NMHV}_{5}$ ect. One
can also obtain the form of $\mbox{NMHV}_{3}$ at tree level using
the representation of $\mbox{NMHV}_{3}$ as MHV vertex in conjugated
$\bar{\eta}$ variables
\cite{Generalized_unitarity_N=4_super-amplitudes}.  We will use both
methods as consistency check and will discuss the structure of
$\overline{\mbox{MHV}}_3=\mbox{NMHV}_{3}$ form factor in the next
sub section at tree and one loop level.

For the triple cut integrand situation is the similar and one have
to consider the following type of integrands:
$$
\mbox{MHV}\times\mbox{MHV}\times\mbox{MHV},
$$
and
$$
\mbox{NMHV}\times\mbox{MHV}\times\overline{\mbox{MHV}}_3.
$$
The last case can be considered in full analogy with the quadruple
cut case. The case of triple MHV cut is more complicated and one
have to use explicit kinematical solutions of
\cite{Triangle_coefficients}. However in the case of $T^{2m}$
triangles at least in the case of $n=3,4$ one can take a short cut
and fix the coefficient using the (\ref{IR_consist}) without any
direct computations.

Let's now discuss the general structure of
$\mbox{MHV}\times\mbox{MHV}\times\mbox{MHV}\times\overline{\mbox{MHV}}_3$
quadruple cuts which correspond in general to the coefficients
$C^{3m}$ before $B^{3m}$ integrals (the coefficients before
$B^{2mh}$ and $B^{1m}$ in general can be obtained using different
combinations of $C^{3m}$ coefficients and cuts involving NMHV
vertexes just as in the amplitude case
\cite{Generalized_unitarity_N=4_super-amplitudes}). There are two
types of configuration of vertexes which we will call 1) and 2) (see
fig.\ref{NMHV_R1} and \ref{NMHV_R2}) and the special limiting case
of 1) (see fig.\ref{NMHV_R1lim}). We will discuss this last case in
details other cases can be considered in the same fashion. The
coefficient $C^{3m}_{s_{r+1...t-1},s_{s...t-1},q^2}$ before
$B^{3m}_{s_{r+1...t-1},s_{t...r-1},q^2}$ scalar box integral is
given by the following expression:
\begin{figure}[h]
 \begin{center}
  \epsfxsize=7cm
 \epsffile{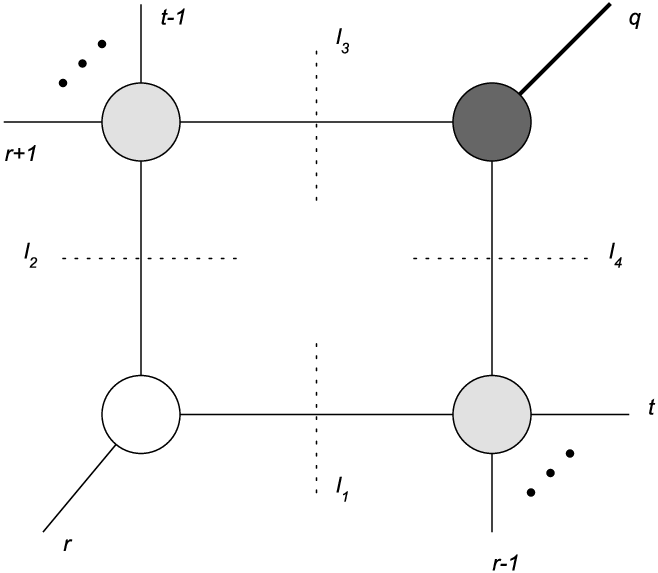}
 \end{center}\vspace{-0.2cm}
 \caption{Diagrammatical representation of the $\tilde{R}^{(1)}_{rtt}$. }\label{NMHV_R1lim}
 \end{figure}
\begin{eqnarray}\label{Definition_of_tilde_R}
C^{3m}_{s_{r+1...t-1},s_{t...r-1},q^2}&=&\int\prod_{i=1}^4d^4\eta^A_{l_{i}}\frac{\delta^4_{GR}(\sum_{-l_4,l_3}\lambda_i\eta_i^{a}+\gamma^{a})\delta^4_{GR}(\sum_{-l_4,l_3}\lambda_i\eta_i^{\dot{a}})}{\langle
l_3l_4\rangle^2}
\nonumber\\
&\times&\frac{\hat{\delta}^4(\eta_r[l_2l_1]+\eta_{l_2}[l_1r]-\eta_{l_1}[rl_2])}{[rl_2][l_2l_1][l_1r]}
\frac{\delta^8(\sum_{i=r+1}^{s-1}\lambda_i\eta^A_i+\lambda_{l_2}\eta_{l_2}^A-\lambda_{l_3}\eta_{l_3}^A)}{\langle
l_3 l_2 \rangle \ldots \langle s-1 l_3
\rangle}\nonumber\\&\times&\frac{\delta^8(\sum_{i=s}^{r-1}\lambda_i\eta^A_i+\lambda_{l_4}\eta_{l_4}^A-\lambda_{l_1}\eta_{l_1}^A)}{\langle
l_1 l_4 \rangle \ldots \langle r-1 l_1 \rangle}.
\end{eqnarray}
This expression can be simplified by using the following strategy
similar to those used in
\cite{Generalized_unitarity_N=4_super-amplitudes}: one has to split
the third and the fourth $\delta^8$ functions into products of two
$\delta_{GR}^4$ (or use (\ref{integration_trick}) prescription) and
add their arguments to the arguments of firs two $\delta_{GR}^4$'s.
After that the dependence on $\eta_{l_i}$ cancels out and we obtain
the $\delta_{GR}^4$'s with the total super momentum as their
argument. The remaining delta functions can be integrated using
(\ref{supersumm_delta_to_hatt_deltas}) which after some
simplifications gives the following result:
\begin{eqnarray}
C^{3m}_{s_{r+1...t-1},s_{t...r-1},q^2}&=&Z_n^{tree,MHV}\frac{\langle
12\rangle \ldots \langle
n1\rangle}{den.}\left(\frac{[l_1r]}{[l_3l_4]}\right)^4\nonumber\\&\times&
\hat{\delta}^4\left(\sum_{i=t}^{r-1}\eta_i\langle
i|l_4l_3|l_1\rangle+\sum_{i=r}^{s-1}\eta_i\langle
i|l_3l_4|l_1\rangle\right),\nonumber\\
\end{eqnarray}
where $den.$ is the product of all denominators in
(\ref{Definition_of_tilde_R}). The argument of the $\hat{\delta}^4$
can be simplified using the kinematical constraints of
$\overline{\mbox{MHV}}_3$ vertex which implies relation
\begin{equation}
\langle i|X|l_1\rangle=\langle
i|X|r\rangle\frac{[rl_2]}{[l_1l_2]},~X=l_4l_3,l_3l_4,
~\lambda_i~\mbox{is arbitrary}.
\end{equation}
So using momentum conservation conditions associated with the MHV
and $\overline{\mbox{MHV}}_3$ vertexes we obtain:
\begin{eqnarray}
C^{3m}_{s_{r+1...t-1},s_{t...r-1},q^2}&=&Z_n^{tree,MHV}\frac{\langle
12\rangle \ldots \langle
n1\rangle}{den.}\left(\frac{[l_1r][l_2r]}{[l_3l_4][l_1l_2]}\right)^4\nonumber\\
&\times&\hat{\delta}^4\left(\sum_{i=t}^{r-1}\eta_i\langle
i|qp_{r...t-1}|r\rangle+\sum_{i=r}^{s-1}\eta_i\langle
i|qp_{t...r-1}|r\rangle
\right),\nonumber\\
\end{eqnarray}
which finally can be written as:
\begin{eqnarray}\label{tildeR_1_deff}
C^{3m}_{s_{r+1...t-1},s_{t...r-1},q^2}&=&Z_n^{tree,MHV}\tilde{R}_{rtt}^{(1)}\frac{1}{2}\Delta_{s_{r+1...t-1},s_{t...r-1},q^2}\nonumber\\
\tilde{R}_{rtt}^{(1)}&=&\frac{\langle
tt-1\rangle\hat{\delta}^4\left(\sum_{i=t}^{r-1}\eta_i\langle
i|qp_{r...t-1}|r\rangle+\sum_{i=r}^{s-1}\eta_i\langle
i|qp_{t...r-1}|r\rangle\right)}{q^4\langle r|p_{r...t-1}q|t\rangle\langle r|p_{t...r}q|t-1\rangle\langle r|x_{t...r-1}q|r\rangle},\nonumber\\
\Delta_{s_{r+1...t-1},s_{t...r-1},q^2}&=&s_{r...t-1}s_{r...t}-s_{t...r-1}s_{r+1...t-1}.
\end{eqnarray}
This expression can be further simplified and written in a compact
form if we introduce dual Grassmann coordinates:
\begin{eqnarray}
\langle\theta_{rt}|=\sum_{i=r}^{t-1}\eta_i\langle i|
=\sum_{i=r}^{t-1}\eta_i\lambda_i,
\end{eqnarray}
\begin{eqnarray}\label{tildeR_1_deff_dualvar}
C^{3m}_{s_{r+1...t-1},s_{t...r-1},q^2}&=&Z_n^{tree,MHV}\tilde{R}_{rtt}^{(1)}\frac{1}{2}\Delta_{s_{r+1...t-1},s_{t...r-1},q^2}\nonumber\\
\tilde{R}_{rtt}^{(1)}&=&\frac{\langle
tt-1\rangle\hat{\delta}^4\left(\langle
\theta_{tr}|x_{1n+1}x_{rt}|r\rangle+\langle
\theta_{rs}|x_{1n+1}x_{tr}|r\rangle\right)}{x_{1n+1}^4\langle r|x_{rt}x_{1n+1}|t\rangle\langle r|x_{tr}x_{1n+1}|t-1\rangle\langle r|x_{tr}x_{1n+1}|r\rangle},\nonumber\\
\Delta_{s_{r+1...t-1},s_{t...r-1},q^2}&=&x_{rt}^2x_{rt+1}^2-x_{tr}^2x_{r+1t}^2.
\end{eqnarray}

Note, that the labels $r,s,t$ when we are using standard helicity
spinor notations belong to the corresponding momenta, and obeys
periodicity conditions $i+n\equiv i,~i\leq n$. But when we use dual
coordinates $x,\theta$ which lives on the infinite periodic contour
there are no periodicity conditions on the $r,s,t$ labels any more.
Throughout this paper if not mentioned otherwise we will think of
$r,s,t$ as labels belonging to the corresponding momenta, and so
obeys periodicity conditions $i+n\equiv i,~i\leq n$.

Similar computations in the $1)$ and $2)$ cases give the results:
\begin{eqnarray}\label{R_1_deff}
C^{3m}_{s_{s-1...r+1}s_{t...r-1}s_{t...s-1}}&=&Z_n^{tree,MHV}R_{rst}^{(1)}
\frac{1}{2}\Delta_{s_{s-1...r+1}s_{t...r-1}s_{t...s-1}},\nonumber\\
R_{rst}^{(1)}&=&\frac{\langle s-1s\rangle\langle
t-1t\rangle\hat{\delta}^4\left(\langle
\theta_{tr}|x_{ts}x_{rs}|r\rangle+\langle
\theta_{rs}|x_{ts}x_{tr}|r\rangle\right)}{x_{ts}^2\langle
r|x_{rs}x_{ts}|t-1\rangle\langle r|x_{rs}x_{ts}|t\rangle\langle
r|x_{tr}x_{ts}|s-1\rangle\langle
r|x_{tr}x_{ts}|s\rangle},\nonumber\\
\Delta_{s_{s-1...r+1}s_{t...r-1}s_{t...s-1}}&=&s_{r...s-1}s_{r...t}-s_{t...r-1}s_{r+1...s-1}
=x_{rs}^2x_{rt+1}^2-x_{tr}^2x_{r+1s}^2,
\end{eqnarray}
\begin{eqnarray}\label{R_2_deff}
C^{3m}_{s_{r...s}s_{t..r-1}s_{t...r-1}}&=&Z_n^{tree,MHV}R_{rst}^{(2)}\frac{1}{2}
\Delta_{s_{r...s}s_{t..r-1}s_{t...r-1}},\nonumber\\
R_{rst}^{(2)}&=&\frac{\langle s-1s\rangle\langle
t-1t\rangle\hat{\delta}^4\left(\langle
\theta_{tr}|x_{st}x_{sr}|r\rangle+\langle
\theta_{rs}|x_{st}x_{tr}|r\rangle\right)}{x_{st}^2\langle
r|x_{sr}x_{st}|t-1\rangle\langle r|x_{sr}x_{st}|t\rangle\langle
r|x_{tr}x_{st}|s-1\rangle\langle
r|x_{tr}x_{st}|s\rangle},\nonumber\\
\Delta_{s_{r...s}s_{t..r-1}s_{t...r-1}}&=&s_{s...r-1}s_{t...r}-s_{r...t-1}s_{s...r}
=x_{sr}^2x_{tr+1}^2-x_{rt}^2x_{sr+1}^2.
\end{eqnarray}
Note that these expressions are very similar to those of the
coefficients $R_{rst}$ in the NMHV amplitudes
\cite{Generalized_unitarity_N=4_super-amplitudes}. The structures of
$R_{rst}^{(1)},~R_{rst}^{(2)}$ and $\tilde{R}_{rtt}^{(1)}$ are in
fact identical and the only difference is the rearrangements of the
sums in $x$ and $|\theta \rangle$ dual coordinates which are made in
such a way that to avoid the dependance on $q$ and $\gamma$ axillary
variables which parameterize the dependence on the (super)momentum
of the operator. Such rearrangements are always possible due to the
total (super)momentum conservation conditions in $Z_n^{tree,MHV}$.

In the case of amplitudes there are large sets of relations between
different combinations of $R_{rst}$
\cite{DualConfInvForAmplitudesCorch}. For example
\begin{equation}
R_{r,r+2,t}=R_{r+2,t,r+1}.
\end{equation} For the case of form
factors one can show that for $n=4$ such relation gives
\begin{equation}
R^{(1)}_{r,r+2,t}=R^{(2)}_{r+2,t,r+1}.
\end{equation} To see this
one has to consider relation $R_{r,r+2,t}=R_{r+2,t,r+1}$ for $n=6$
and then put $q_5^a=\lambda^{'}\eta^{'a}$,
$q_6^a=\lambda^{''}\eta^{''a}$ and $q_5^{\dot{a}}=q_6^{\dot{a}}=0$.
The clockwise cyclic order of external legs is implied. Such
relations in the case of amplitudes can be easily proved using
momentum twistor representation \cite{Twistors}. It is interesting
to note that obtained here $R^{(i)}_{rst}$ and
$\tilde{R}_{rtt}^{(1)}$ coefficients can be rewritten in momentum
twistor notations as well, and are equal to special cases of
$[abcde]$ momentum twistor invariants \cite{Twistors}. We are going
to discuss this in more details in separate publication.

In the NMHV computations one will also encounter the
$T^{3m}_{s_{r...s},s_{s+1...t},q^2}$ three mass triangles. The
coefficients $C^{3m}_{s_{r...s},s_{s+1...t},q^2}$ before such
triangles should be fixed by the triple MHV cuts. In such case the
explicit solutions for $\lambda_{l_i}$ and
$l_i^{\alpha\dot{\alpha}}$ should be used, so it is problematic to
obtain representation for such coefficients in terms of only
$\lambda_i,\tilde{\lambda}_i$ spinors, which corresponds to external
momenta. For the integrand of such cut one has:
\begin{eqnarray}
&&\int\prod_{i=1}^3d^4\eta^A_{l_{i}}
\frac{\delta^4_{GR}(\sum_{l_2,l_3}\lambda_i\eta_i^{a}+\gamma^{a})\delta^4_{GR}(\sum_{l_2,l_3}\lambda_i\eta_i^{\dot{a}})}{\langle
l_2l_3\rangle^2}\frac{\delta^8(\sum_{i=r}^{s}\lambda_i\eta^A_i+\lambda_{l_1}\eta_{l_1}^A-\lambda_{l_2}\eta_{l_2}^A)}{\langle
l_1 l_2 \rangle \ldots \langle s l_1
\rangle}\nonumber\\&\times&\frac{\delta^8(\sum_{i=s+1}^{t}\lambda_i\eta^A_i+\lambda_{l_3}\eta_{l_3}^A-\lambda_{l_1}\eta_{l_1}^A)}{\langle
l_1 l_3 \rangle \ldots \langle s+1 l_1 \rangle}.
\end{eqnarray}
After the integration over Grassamann variables one obtains:
\begin{eqnarray}
&&Z^{MHV,tree}_n\frac{\langle rt\rangle\langle ss+1
\rangle\hat{\delta}^4\left(\sum_{i=r}^s\eta_{i}\langle
l_1l_3\rangle\langle l_2i\rangle+\sum_{i=s+1}^t\eta_i\langle l_il_2
\rangle\langle l_3i \rangle\right)}{\langle rl_2\rangle\langle
l_2l_1\rangle\langle l_1s\rangle\langle s+1l_1\rangle\langle
l_1l_3\rangle \langle l_3t \rangle \langle l_1l_2\rangle^2}.
\end{eqnarray}
\begin{figure}[h]
 \begin{center}
  \epsfxsize=7cm
 \epsffile{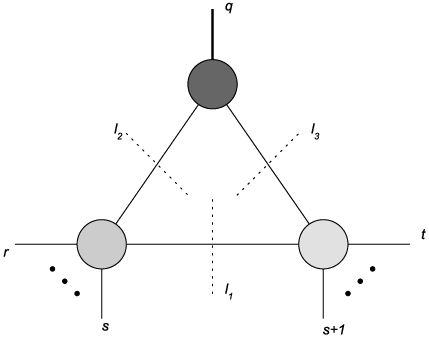}
 \end{center}\vspace{-0.2cm}
 \caption{Diagrammatical representation of triple MHV cut. }\label{Definition_of_R_for_triple_cut}
 \end{figure}
At this moment one has to use explicit solutions for
$\lambda_{l_i}$. After substitution of these solutions one can take
$t\rightarrow \infty$ limit and obtain:
\begin{eqnarray}
C^{3m}_{s_{r...s},s_{s+1...t},q^2}&=&Z^{MHV,tree}_n\frac{1}{2}\mathcal{R}_{rst}\Delta_{s_{r...s},s_{s+1...t},q^2},\nonumber\\
\mathcal{R}_{rst}&=&\sum_{\pm S}\frac{\gamma
}{K_1^2K_2^2}\left(\frac{K_1^2}{\gamma}-1\right)^{-3}\frac{\langle
rt\rangle\langle ss+1
\rangle}{\prod_{i=1}^{n}\langle iK_1^{\flat}\rangle}\nonumber\\
&\times&\hat{\delta}^4\left(K_1^2/\gamma\sum_{i=r}^s\eta_{i}\langle
K_1^{\flat}i\rangle+(K_1^2/\gamma-1)\sum_{i=s+1}^t\eta_i\langle
K_1^{\flat}i \rangle\right),\nonumber\\
\Delta_{s_{r...s},s_{s+1...t},q^2}&=&q^2=x_{1n+1}^2.
\end{eqnarray}
Here $K_1=p_{s+1...t}=x_{s+1...t-1},~K_2=q=x_{1n+1}$ and
$K_i^{\flat}$ - are the massless projections of one massive leg in
the direction of another masslessly projected leg (see appendix).

Let's now discuss the beginning of the recursive procedure discussed
above for the NMHV tree level form factors which one will need in
general for computation of quadruple cuts. We will discuss one step
of this procedure and obtain answers for $n=3$ and $n=4$ point NMHV
form factors at tree and one loop level.

\subsection{3 point NMHV form factor at tree and one loop level}
\begin{figure}[t]
 \begin{center}
  \epsfxsize=11cm
 \epsffile{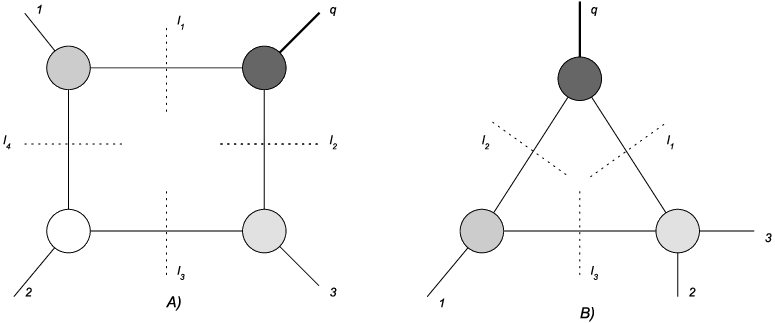}
 \end{center}\vspace{-0.2cm}
 \caption{Contributing cuts for the box $B^{1m}$ and triangle $T^{2m}$ scalar integrals coefficients to
 $\mbox{NMHV}_3$
  form factor. Permutations of external momenta are not shown.}\label{NMHV_quadruple_cuts_3}
 \end{figure}
Let's consider the representation for $\mbox{MHV}_3$ form factor at
tree level in terms of $\bar{\eta}$ variables
\cite{Generalized_unitarity_N=4_super-amplitudes}:
\begin{equation}
\langle\overline{\Omega_3}|\overline{T}^{(0)}_{\dot{a}\dot{b}}|0\rangle^{MHV}=
\overline{F}_3^{MHV}(\{\lambda,\tilde{\lambda},\bar{\eta}\})=
\frac{\delta^4_{GR}
\left(\sum_{i=1}^{3}\tilde{\lambda}_{\alpha}^{i}\bar{\eta}^{a}_i\right)}{[12][23][31]},
\end{equation}
where
\begin{equation}
\overline{T}^{(0)\dot{a}\dot{b}}=(\overline{\mathcal{T}^{ab}})^2\Big|_{\theta=\bar{\theta}=0}
=(\overline{\phi^{ab}})^2=(\phi_{ab})^2=(\phi^{\dot{a}\dot{b}})^2.
\end{equation}
We can perform the Fourier transformation from $\bar{\eta}$ to
$\eta$'s \cite{Generalized_unitarity_N=4_super-amplitudes} and write
representation $F^{\overline{MHV}}_3$ for $\overline{\mbox{MHV}}_3$
form factor at tree level in terms of $\eta$'s as:
\begin{equation}
F^{\overline{MHV}}_3(\{\lambda,\tilde{\lambda},\eta\})=\int\prod_{i=1}^3d^4\overline{\eta}_i~exp(\eta^A_i\bar{\eta}^i_A)
\overline{F}_3^{MHV}(\{\lambda,\tilde{\lambda},\bar{\eta}\}).
\end{equation}
This integral can be evaluated exactly and for
$Z^{\overline{MHV}}_3$ which is connected with
$F^{\overline{MHV}}_3$ as
\begin{eqnarray}
\mathcal{F}_n^{tree,\overline{MHV}}&=&\exp(\theta_{a}^{\alpha}q^{a}_{\alpha})F_n^{tree,\overline{MHV}}\nonumber\\
Z^{tree,\overline{MHV}}_n&=&\hat{T}[\mathcal{F}_n^{tree,\overline{MHV}}],
\end{eqnarray}
we can obtain, noticing that $\overline{\mbox{MHV}}_3$ form factor
is also $\mbox{NMHV}_3$ one:
\begin{equation}\label{NMHV_3_tree_simple}
Z^{tree,\overline{MHV}}_3=Z^{tree,NMHV}_3=\delta^4_{GR}(\sum_{i=1}^{n}\lambda_{\alpha}^{i}\eta^{a}_i
+
\gamma^a_{\alpha})\prod_{i=1}^3\hat{\delta}^2(\eta^{\dot{a}}_i)\frac{\hat{\delta}^2(\eta_1^a[23]+\mbox{cycl.perm.})}{[12][23][31]}.
\end{equation} This expression is explicitly cyclically
invariant. Using identities:
\begin{equation}
\langle\theta_{12}|x_{14}x_{34}|2\rangle+\langle
\theta_{24}|x_{14}x_{12}|2\rangle=\langle12\rangle\langle23\rangle(\eta_1[23]+\eta_2[31]+\eta_3[12]),
\end{equation}
and
\begin{equation}
x_{14}^4\prod_{i=1}^3\hat{\delta}^2(\eta^{\dot{a}}_i)=\delta^4_{GR}(\sum_{i=1}^3\lambda_i\eta^{\dot{a}}_i)
\hat{\delta}^2(\eta_1^{\dot{a}}[23]+\eta_2^{\dot{a}}[31]+\eta_3^{\dot{a}}[12]),
\end{equation}
one can write $Z^{tree,NMHV}_3$ as:
\begin{equation}
Z^{tree,NMHV}_3=Z^{tree,MHV}_3\tilde{R}_{211}^{(1)},
\end{equation}
where $\tilde{R}_{211}^{(1)}$ is given by (\ref{tildeR_1_deff}) for
$n=3$. Note also that in this case ($n=3$) the following identity
holds:
$\tilde{R}_{211}^{(1)}=\tilde{R}_{322}^{(1)}=\tilde{R}_{133}^{(1)}$
which is consequence of the cyclical symmetry of the initial
expression for NMHV three point form factor at tree level. Using
this we can write $Z^{tree,NMHV}_3$ in manifestly cyclically
invariant form using $\tilde{R}^{(1)}_{rtt}$ coefficients:
\begin{equation}
Z^{tree,NMHV}_3=Z^{MHV,tree}_3\frac{1}{3}(1+\mathbb{P}+\mathbb{P}^2)\tilde{R}^{(1)}_{211},
\end{equation}
where  $\mathbb{P}$ is permutation operator. We can now use
$Z^{tree,NMHV}_3$ in the computations of $Z^{(1),NMHV}_4$, but
before that let's perform one loop computation also for the
$Z^{(1),NMHV}_3$, which is essentially trivial. The contributing
scalar integrals are the same as in the $\mbox{MHV}_3$ case:
$B^{1m}$ and $T^{2m}$. Moreover because
$\mbox{NMHV}_3=\overline{\mbox{MHV}}_3$ the ratio of the one loop
correction over tree result will be the same as in $\mbox{MHV}_3$.
The coefficients $C^{1m}$ before $B^{1m}$ integrals are fixed by the
quadruple cuts of the type (\ref{tildeR_1_deff}). For example for
$B^{1m}(1,2,3|q^2)$ integral we obtain:
$$
C^{1m}_{q^2}=Z^{tree,MHV}_3\frac{1}{2}\tilde{R}_{211}^{(1)}x^2_{24}x^2_{13}.
$$
The computation of the $C^{2m}$ coefficients before scalar integrals
$T^{2m}$ is more complicated. The integrand of corresponding cut is
given by the triple product of MHV vertexes, and there are not
enough kinematical conditions to express $l_i^{\alpha\dot{\alpha}}$
momenta and $\lambda_{l_i}$ spinors in terms of the external momenta
and spinors associated with them not using exact form of $\pm S$
solutions. However there is only one type of such coefficients and
one can fix their value by requiring the (\ref{IR_consist})
condition must be satisfied. We will use this approach from now on.

One can see that in the case under consideration the
(\ref{IR_consist}) condition satisfied if and only if (let's
consider coefficient before $T^{2m}(1|q^2,(2+3)^2)$ integral):
$$
C^{2m}_{s_{23},q^2}=Z^{tree,MHV}_3\frac{1}{2}\tilde{R}_{211}^{(1)}(x^2_{24}-x^2_{14}).
$$

Coefficients before other integrals can be obtained by action of
permutation operator $\mathbb{P}$. Combining all contributions
together we can arrange the final result as:
\begin{eqnarray}
Z_{3}^{(1),NMHV}/Z_{3}^{tree,NMHV}&=&\frac{1}{2}\mathcal{B}^{1m}(1,2,3|q^2)+\frac{1}{2}\mathcal{B}^{1m}(1,3,2|q^2)+\frac{1}{2}\mathcal{B}^{1m}(2,1,3|q^2)
\nonumber\\
&+&\mathcal{T}^{2m}(1|q^2,(2+3)^2)+
\mathcal{T}^{2m}(2|q^2,(1+3)^2) \nonumber\\
&+&\mathcal{T}^{2m}(3|q^2,(1+2)^2).
\end{eqnarray}
Substituting the expansions in $\epsilon$ of $\mathcal{B}^{1m}$ and
$\mathcal{T}^{2m}$ functions we obtain the
(\ref{MHV_3_fin_part_Dual_var}) result.

\subsection{4 point NMHV form factor at one loop}
\begin{figure}[t]
 \begin{center}
 \leavevmode
  \epsfxsize=11cm
 \epsffile{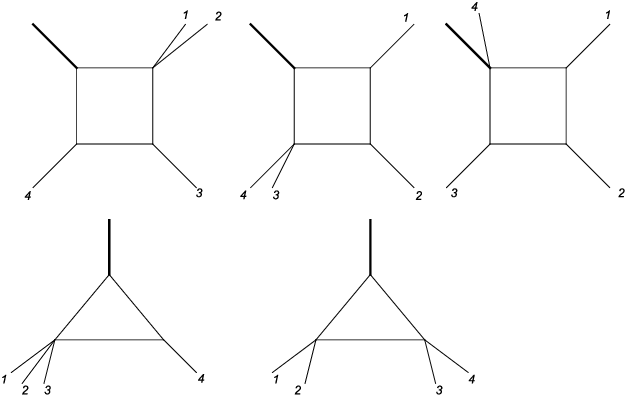}
 \end{center}\vspace{-0.2cm}
 \caption{List of all contributing cyclically inequivalent scalar integrals to $\mbox{NMHV}_4$}\label{fig2}
 \end{figure}
Now we are ready for the computation of $\mbox{NMHV}_4$ form factor
at one loop. The contributing scalar integrals in this case
are\footnote{The $B^{2me}$ scalar boxes are absent by the same
reasons as in the 6 point NMHV amplitude
\cite{Generalized_unitarity_N=4_super-amplitudes}.}
$B^{2mh},~B^{1m},T^{2m}$ and $T^{3m}$. The first three types of
integrals are IR divergent, while the last one (tree mass triangle)
is IR finite. Let us remind the reader that there are no other
contributing scalar triangle integrals for the reasons discussed in
the previous section. Let's concentrate first on the IR divergent
part of the answer. We will label the coefficients before cyclically
inequivalent type of integrals as:
\begin{eqnarray}
&&C^{2mh}_{q^2,(1+2)^2},~C^{2mh}_{q^2,(4+3)^2},~C^{1m}_{(q-4)^2},C^{2m}_{q^2,(1+2+3)^2}.
\end{eqnarray}

For $C^{2mh}_{q^2,(1+2)^2}$ we have two types of cuts contributing
(see fig.\ref{fig_NMHV4_quadr_triple_cuts}):
\begin{equation}
C^{2mh}_{q^2,(1+2)^2}=A)+B).
\end{equation}
These cuts give us:
\begin{equation}
C^{2mh}_{q^2,(1+2)^2}=Z^{tree,MHV}_4\frac{1}{2}(\tilde{R}^{(1)}_{311}+R^{(2)}_{413})x^2_{35}x^2_{14}.
\end{equation}

For $C^{1m}_{(q-4)^2}$ we have two types of contributions:
\begin{equation}
C^{1m}_{(q-4)^2}=C)+D).
\end{equation}
These cuts give us:
\begin{equation}
C^{1m}_{(q-4)^2}=Z^{tree,MHV}_4\frac{1}{2}(\tilde{R}^{(1)}_{144}+R^{(1)}_{241})x^2_{24}x^2_{13}.
\end{equation}
Note that $C)$ cut contains vertex which is $\mbox{NMHV}$ 3-point
form factor. To evaluate such cuts we use the same technique as in
\cite{Generalized_unitarity_N=4_super-amplitudes}: we substitute
explicit expression for $\mbox{NMHV}$ 3-point form factor and then
use the kinematical constraints attached to the neighboring
$\overline{\mbox{MHV}}_3$ vertexes which imply
$\lambda_{l_2}\sim\lambda_1$ and $\lambda_{l_3}\sim\lambda_3$. Using
the cyclical symmetry of $\mbox{NMHV}$ 3-point form factor we can
obtain the following relations for $\tilde{R}^{(1)}$:
\begin{equation}\label{relations_for_tildeR}
\tilde{R}^{(1)}_{144}=\tilde{R}^{(1)}_{311},\tilde{R}^{(1)}_{244}=\tilde{R}^{(1)}_{211}.
\end{equation}

For $C^{2mh}_{q^2,(4+3)^2}$ similar to $C^{2mh}_{q^2,(1+2)^2}$ we
have:
\begin{equation}
C^{2mh}_{q^2,(4+3)^2}=Z^{tree,MHV}_4\frac{1}{2}(\tilde{R}^{(1)}_{244}+R^{(2)}_{142})x^2_{13}x^2_{25}.
\end{equation}

For $C^{2m}_{q^2,(1+2+3)^2}$ we have two contributing cuts. Cut
involving NMHV 5 point tree amplitude can be evaluated using
explicit expression for NMHV 5 point tree amplitude and kinematical
constraints associated with the neighboring
$\overline{\mbox{MHV}}_3$ vertex, which imply
$\lambda_{l_2}\sim\lambda_{1}$. We will adjust the value of the
other triple MHV cut using the (\ref{IR_consist}) just as in the
previous $\mbox{NMHV}_3$ case. Finally we have:
\begin{equation}
C^{T2m}_{q^2,(1+2+3)^2}=Z^{tree,MHV}_4\frac{1}{2}(\tilde{R}^{(1)}_{311}+R^{(2)}_{413})(x^2_{14}-x^2_{15}).
\end{equation}
One can notice that
$C^{2m}_{q^2,(4+2+3)^2}=\mathbb{P}C^{2m}_{q^2,(1+2+3)^2}$, so we can
write all contributions from the IR divergent triangle integrals in
the following form:
\begin{equation}
C^{2m}_{q^2,(1+2+3)^2}\mathcal{T}^{2m}_{q^2,(1+2+3)^2}+C^{2m}_{q^2,(4+2+3)^2}\mathcal{T}^{2m}_{q^2,(4+2+3)^2}+perm.
=2C^{2m}_{q^2,(1+2+3)^2}\mathcal{T}^{2m}_{q^2,(1+2+3)^2}+perm.
\end{equation}

So, finally gathering all IR divergent contributions together we
obtain
\begin{eqnarray}
Z_4^{NMHV,(1)}/Z_4^{tree,MHV}\Big|_{IR}&=&\frac{1}{2}(\tilde{R}^{(1)}_{311}+R^{(2)}_{413})\mathcal{B}^{2mh}(3,4|q^2,(1+2)^2)\Big|_{IR}
\nonumber\\
&+&\frac{1}{2}(\tilde{R}^{(1)}_{144}+R^{(1)}_{241})\mathcal{B}^{1m}(1,2,3|(q-4)^2)\Big|_{IR}
\nonumber\\
&+&\frac{1}{2}(\tilde{R}^{(1)}_{244}+R^{(2)}_{142})\mathcal{B}^{2mh}(1,2|q^2,(3+4)^2)\Big|_{IR}
\nonumber\\
&+&\frac{1}{2}(\tilde{R}^{(1)}_{311}+R^{(2)}_{413})2\mathcal{T}^{2m}(4|q^2,(1+2+3)^2)\Big|_{IR}+perm.\nonumber\\
\end{eqnarray}
Using equation
\begin{equation}
R^{(2)}_{413}=R^{(1)}_{241}
\end{equation}
which is consequence of $R_{r,r+2,t}=R_{r+2,t,r+1}$ relation for
amplitudes, eq.(\ref{relations_for_tildeR}) and
$$
\tilde{R}^{(1)}_{244}=\tilde{R}^{(1)}_{211}=\mathbb{P}\tilde{R}^{(1)}_{144},
$$
also noticing that $R^{(1)}_{312}=\mathbb{P}R^{(1)}_{241}$ we can
write:
\begin{eqnarray}
Z_4^{NMHV,(1)}/Z_4^{tree,MHV}\Big|_{IR}&=&\frac{1}{2}(\tilde{R}^{(1)}_{311}+R^{(1)}_{241})\{\mathcal{B}^{2mh}(3,4|q^2,(1+2)^2)
\nonumber\\
&+&\mathcal{\mathcal{B}}^{2mh}(4,1|q^2,(3+2)^2)+\mathcal{B}^{1m}(1,2,3|(q-4)^2)
\nonumber\\&+&2\mathcal{T}^{2m}(4|q^2,(1+2+3)^2)\}\Big|_{IR}+perm.
\nonumber\\
\end{eqnarray} Using explicit
expressions for the IR divergent parts of integrals
\begin{eqnarray}
\mathcal{B}^{2mh}(3,4|q^2,(1+2)^2)\Big|_{IR}&=&\frac{\mu^{-2\epsilon}}{\epsilon^2}\left(s_{43}^{\epsilon}+2(4-q)^{2\epsilon}-q^{2\epsilon}-s_{12}^{\epsilon}\right),\nonumber\\
\mathcal{B}^{2mh}(4,1|q^2,(3+2)^2)\Big|_{IR}&=&\frac{\mu^{-2\epsilon}}{\epsilon^2}\left(s_{14}^{\epsilon}+2(4-q)^{2\epsilon}-q^{2\epsilon}-s_{23}^{\epsilon}\right),\nonumber\\
\mathcal{B}^{1m}(1,2,3|(q-4)^2)\Big|_{IR}&=&\frac{\mu^{-2\epsilon}}{\epsilon^2}\left(2s_{32}^{\epsilon}+2s_{21}^{\epsilon}-2(4-q)^{2\epsilon}\right),\nonumber\\
\mathcal{T}^{2m}(4|q^2,(1+2+3)^2)\Big|_{IR}&=&\frac{\mu^{-2\epsilon}}{\epsilon^2}\left(q^{2\epsilon}-(4-q)^{2\epsilon}\right),
\end{eqnarray}
one can see that IR divergent part of the NMHV four point form
factor indeed has the form:
\begin{equation}
Z_4^{NMHV,(1)}/Z_4^{tree,MHV}\Big|_{IR}=\frac{1}{2}
(1+\mathbb{P}+\mathbb{P}^2+\mathbb{P}^3)(\tilde{R}^{(1)}_{311}+R^{(1)}_{241})\sum_{i=1}^4\frac{1}{\epsilon^2}\left(\frac{s_{ii+1}}{\mu^2}\right)^{\epsilon}.
\end{equation}
From this we conclude that
\begin{equation}\label{NMHV4_tree}
Z_4^{NMHV,tree} =Z_4^{MHV,tree}\frac{1}{2}
(1+\mathbb{P}+\mathbb{P}^2+\mathbb{P}^3)(\tilde{R}^{(1)}_{311}+R^{(1)}_{241}).
\end{equation}

Let's now consider the contribution of IR finite
$T^{3m}((1+2)^2,(3+4)^2,q^2)$ scalar integral. For the corresponding
coefficient $C^{3m}_{(1+2)^2,(3+4)^2,q^2}$ using previous results we
immediately obtain:
\begin{equation}
C^{3m}_{(1+2)^2,(3+4)^2,q^2}=Z^{MHV,tree}\mathcal{R}_{124}q^2.
\end{equation}
The finite part of four point NMHV form factor then can be written
in the following form:
\begin{eqnarray}
Z_4^{NMHV,(1)}/Z_4^{tree,MHV}\Big|_{fin}&=&(1+\mathbb{P}+\mathbb{P}^2+\mathbb{P}^3)\frac{1}{2}(\tilde{R}^{(1)}_{311}+R^{(1)}_{241})V_4\nonumber\\
&+&(1+\mathbb{P}+\mathbb{P}^2+\mathbb{P}^3)\mathcal{R}_{124}W_4.
\end{eqnarray}
Where ($\mbox{Log}(x)\equiv L(x)$):
\begin{eqnarray}
V_4&=&-2\sum_{i=1}^2\left(\mbox{Li}_2\left(1-\frac{x^2_{ii+2}}{x^2_{14}}\right)+\mbox{Li}_2\left(1-\frac{x^2_{14}}{x^2_{ii+2}}\right)\right)-4\mbox{Li}_2\left(1-\frac{x^2_{15}}{x^2_{14}}\right)\nonumber\\
&+&\sum_{i=1}^2\left(-L^2\left(\frac{x^2_{ii+2}}{x^2_{14}}\right)+
L\left(\frac{x^2_{ii+2}}{x^2_{i+2i+4}}\right)L\left(\frac{x^2_{15}}{x^2_{i+2i+4}}\right)\right)-
L^2\left(\frac{x^2_{13}}{x^2_{24}}\right)-\frac{\pi^2}{3},\nonumber\\
\end{eqnarray}
while $W_4$ is given by Davydychev function \cite{Usyukina:1992jd},
which in our case has the form:
\begin{eqnarray}
W_4&=&\frac{1}{Q}\left(2\mbox{Li}_2(-xR)+2\mbox{Li}_2(-yR)+L(xR)L(yR)+
L\left(\frac{y}{x}\right)L\left(\frac{1+yR}{1+xR}\right)+\frac{\pi^2}{3}\right),\nonumber\\
Q&=&\left((1-x-y)^2-4xy\right)^{1/2},~R=2(1-x-y+R)^{-1},~x=\frac{x^2_{13}}{x^2_{15}},~y=\frac{x^2_{35}}{x^2_{15}}.
\end{eqnarray}
This is the end of computation of the four point NMHV form factor.
Using results obtained in previous section for MHV sector one can in
principle define analog of $R_n^{NMHV}$ ratio function as in the
amplitude case \cite{Generalized_unitarity_N=4_super-amplitudes}.
One also can use (\ref{NMHV4_tree}) as an input in the computation
of the five point NMHV form factor. The structure of such
computation should be essentially similar to the four point case.
The IR divergent part of the answer will be determined by the IR
divergent parts of box and two mass triangle scalar integrals, while
in the finite part there will be additional contributions from IR
finite three mass integrals.

It is interesting to note that the number of $R_{rst}^{(i)}$
coefficients in tree level 3 and 4 point form factors exactly
matches the number of diagrams which one will have for the
corresponding super Wilson loop \cite{Huot} if one will use the
prescriptions of \cite{HarmonyofFF} about selection of diagrams. It
will be interesting to clarify by explicit computations the status
of the Wilson loop/form factor duality in the none-MHV case.

In \cite{Sasha,BKV_SuperForm} it was observed that the following
relation between form factors and amplitudes likely holds
\begin{equation}\label{cojectureAmpl-FF}
Z_n(\{\lambda,\tilde{\lambda},\eta\},0,\{0\})=\hat{T}[\mathcal{F}_n^{MHV}]
(\{\lambda,\tilde{\lambda},\eta\},0,0) = g\frac{\partial
\mathcal{A}_n(\lambda,\tilde{\lambda},\eta)}{\partial g}.
\end{equation}
This relation was verified at tree and one loop level for the MHV
sector. While for the component answer for the 4 point NMHV form
factor the limit $q \rightarrow 0$ was in general singular. If one
considers the coefficients before corresponding scalar integrals in
the case of five point form factor at one loop one can observe the
following: in the limit $q \rightarrow 0,~\gamma^{a}_{\alpha}
\rightarrow 0$ the coefficients before triangle integrals vanish and
so do most of the coefficients before box scalar integrals. The only
coefficients that survive are the coefficients before $B^{1m}$ box
integrals which in this limit reproduce the coefficients for the
five point NMHV amplitude. It is likely the general pattern for the
n point case. One may think that relation (\ref{cojectureAmpl-FF})
holds for such numbers of external legs for which the objects on
both sides of the equality exists. I.e. equality should not hold for
example for the NMHV sector for $n=4$ because there are no $n=4$
NMHV amplitude.


\section{Conclusions}
In this paper the systematic study of form factors in the
$\mathcal{N}=4$ SYM theory in on-shell momentum superspace formalism
is performed for the MHV and NMHV sectors at one loop order of PT by
means of generalized unitarity technique. The use of $\mathcal{N}=4$
covariant methods allows us to obtain answers for any type of
operator from stress tensor multiplet and arbitrary external states
in these sectors. The explicit answers for the 3 and 4 point NMHV
form factors were obtained as well as the n point situation was
discussed. As the byproduct of this investigation the representation
for 3 and 4 point form factors that does not depend on any kind of
"reference spinors" at tree level were obtained.

The application of the generalized unitarity methods to form factors
clarifies several issues: the structure of the basis of scalar
integrals at one loop and the relations between form factors with
operator insertion with zero momenta and the amplitudes.

The recent studies of the structure of the amplitudes and their
relations to Wilson loops in $\mathcal{N}=4$ SYM led to the
formulation of the equation which at least in principle should
define full S-matrix of the theory \cite{HuotEquation,Twistors2DescentEquation} 
at any value of the coupling constant. The conjectured derivation of this
equation is based on the amplitudes/Wilson loops duality. There is
also the conjecture \cite{FormFactorMHV_component_Brandhuber} that
similar duality between form factors and Wilson loops also holds. It
is interesting to investigate whether such duality survives for the
NMHV and other sectors in some form and if it is possible to
formulate similar equation for the form factors. The results
obtained in this paper may be considered as starting point in such
investigation.

\section*{Acknowledgements}

The author would like to thank D. I. Kazakov, G. S. Vartanov for
valuable and stimulating discussions and early collaboration. The
author would like also to thank A. V. Zhiboedov, A. A. Gorsky for
useful comments and Sunyoung Shin for interesting discussions.
Financial support from the Dynasty Foundation, RFBR grant \#
11-02-01177 and the Ministry of Education and Science of the Russian
Federation grant \# 1027.2008.2 is kindly acknowledged.

\appendix

\section{$\mathcal{N}=4$ harmonic superspace}
We  discuss here the reformulation of (\ref{T[superFormfactor]}) in
the $\mathcal{N}=4$ harmonic superspace. Our discussion is based
mostly on section 3 of \cite{SuperCor1}. The $\mathcal{N}=4$
harmonic superspace is obtained by adding additional bosonic
coordinates (harmonic variables) to the $\mathcal{N}=4$ coordinate
superspace or on-shell momentum superspace. These additional bosonic
coordinates parameterize the coset
$$
\frac{SU(4)}{SU(2) \times SU(2)' \times U(1)}
$$
and carry the  $SU(4)$ index $A$, two copies of $SU(2)$ indices $a,
\dot{a}$ and $U(1)$ charge $\pm$
$$
(u^{+a}_{A},~u^{-\dot{a}}_A).
$$

Using these variables one presents all the Grassmannian objects with
$SU(4)_R$ indices. For example, for Grassmannian coordinates in the
original $\mathcal{N}=4$ coordinate superspace
\begin{equation}
\theta^{+a}_{\alpha}=u^{+a}_{A}\theta^A_{\alpha},~
\theta^{-\dot{a}}_{\alpha}=u^{-\dot{a}}_{A}\theta^A_{\alpha},
\end{equation}
and in the opposite direction
\begin{equation}
\theta^A_{\alpha}=\theta^{+a}_{\alpha}u_{+a}^{A}+\theta^{-\dot{a}}_{\alpha}\bar{u}_{+a}^{A}.
\end{equation}

The same can be done with supercharges etc.. Note that harmonic
variable projection leaves helicity properties of the objects
unmodified. Also, similar projections can be performed for
Grassmannian coordinates $\eta^A$ and supercharges
$q_{\alpha}^A,~\bar{q}_{\dot{\alpha}A}$ of on-shell momentum
superspace.

So the $\mathcal{N}=4$ harmonic superspace is parameterized with the
following set of coordinates
\begin{eqnarray}
\mbox{$\mathcal{N}=4$ harmonic
superspace}&=&\{x^{\alpha\dot{\alpha}},
~\theta^{+a}_{\alpha},\theta^{-\dot{a}}_{\alpha},
~\bar{\theta}_{\dot{\alpha}}^{+a},\bar{\theta}_{\dot{\alpha}}^{-\dot{a}},u
\}\nonumber\\
\mbox{or}&&\{\lambda_{\alpha},\tilde{\lambda}_{\dot{\alpha}},~\eta^{+a},\eta^{-\dot{a}},~u\}.
\end{eqnarray}
Using $u$ harmonic variables one can project the $W^{AB}$ superfield
as
$$
W^{AB}\rightarrow W^{AB}u^{+a}_{A}u^{+b}_{B}=\epsilon^{ab}W^{++},
$$
where $\epsilon^{ab}$ is an $SU(2)$ totally antisymmetric tensor and
the Grassmannian analyticity conditions \cite{SuperCor1} such
that\footnote{Strictly speaking this is true only in the free theory
($g=0$), in the interacting theory one has to replace
$D_{\alpha}^A,\bar{D}_{\dot{\alpha}}^A$ by their gauge covariant
analogs, which contain superconnection, but the final result is the
same \cite{SuperCor1}.}
$$
D^{\alpha}_{-\dot{a}}W^{++}=0,~ \bar{D}^{\dot{\alpha}}_{+a}W^{++}=0.
$$

Thus, the superfield $W^{++}$ contains the dependence on half of the
Grassmannian variables $\theta$'s and $\bar{\theta}$'s.
$$
W^{++}=W^{++}(x, ~\theta^{+a}_{\alpha},
\bar{\theta}_{\dot{\alpha}}^{-\dot{a}},u),
$$

Performing the expansion of $W^{++}$   in $u$ \emph{all} the
projections like (\ref{prrojection2}) in $SU(4)_R$ covariant fashion
can be obtained. This is the main purpose of introduction of the
harmonic superspace. The component expansion of $W^{++}$ in
$\theta$'s and $\bar{\theta}$'s can be found in \cite{SuperCor1}.
The lowest component of the $W^{++}$ expansion is
$$
W^{++}(x,0,0,u)=\phi^{++},~\phi^{++}=\frac{1}{2}\epsilon_{ab}u^{+a}_{A}u^{+b}_{B}\phi^{AB},
$$
where according to \cite{SuperCor1}
\begin{equation}\label{QS++=0}
Q^{\alpha}_{-\dot{a}}\phi^{++}=0.
\end{equation}

Using this condition and the translation invariance we can write the
expression for MHV
 "super state -super form factor" at tree level in harmonic superspace \cite{BKV_SuperForm}:
\begin{equation}
\hat{Z}_n^{tree,MHV}(\{\lambda,\tilde{\lambda},\eta\},q,u,\gamma^{+a}_{\alpha})=
\frac{\delta^{+4}(q^{a}_{\alpha}+\gamma^{a}_{\alpha})\delta^{-4}(q^{\dot{a}}_{\alpha})}{\langle
12 \rangle \ldots \langle n1 \rangle},
\end{equation}
where $\delta^{\pm4}$ is the Grassmannian delta function;
$\delta^{\pm4}$ are defined as
\begin{equation}
\delta^{\pm4}(q^{a/\dot{a}}_{\alpha})=\sum_{i,j=1}^{n}\prod_{a/\dot{a},b/\dot{b}=1}^2\langle
ij \rangle \eta^{\pm a/\dot{a}}_i\eta^{\pm b/\dot{b}}_j.
\end{equation}
We can also define $\hat{\delta}^{\pm2}$ as usual Grassmann delta
functions:
\begin{equation}
\hat{\delta}^{\pm2}(X^{a/\dot{a}})=\prod_{a/\dot{a},b/\dot{b}=1}^2X^{a/\dot{a}}.
\end{equation}
The obtained expression for form factor looks just like
(\ref{T[superFormfactor]}), but now both the Grassmannian delta
functions $\delta^{\pm4}$  are $SU(4)_R$ covariant. One can write
also the MHV part of a superamplitude  in a similar manner.
Projecting the condition of superamplitude invariance under
$q^{A}_{\alpha}$ supersymmetry transformations we have
$$
q^{A}_{\alpha}\hat{\mathcal{A}}_n^{tree,MHV}=0\rightarrow
(q^{+a}_{\alpha}+q^{-\dot{a}}_{\alpha})
\hat{\mathcal{A}}_n^{tree,MHV}=0,
$$
and taking into account that the helicity properties of projected
supercharges are not modified we get
\begin{equation}
\hat{\mathcal{A}}_n^{tree,MHV}=
\frac{\delta^{+4}(q^{a}_{\alpha})\delta^{-4}(q^{\dot{a}}_{\alpha})}{\langle
12 \rangle \ldots \langle n1 \rangle}.
\end{equation}
Now both $\hat{\mathcal{A}}_n^{tree,MHV}$ and $\hat{Z}^{tree,MHV}_n$
are $SU(4)_R$ invariant and one can use them in unitarity based
computations, where Grassmann integration (super summation) should
be performed separately for $d^{+2}\eta$ and $d^{-2}\eta$.

We see now that all results obtained in previous sections can be
simply generalized to $SU(4)_R$ covariant harmonic superspace
version. We have to replace common $\hat{Z}^{tree,MHV}_n$ prefactor
by its harmonic superspace generalization and replace all
$\hat{\delta}^4(X^{A}_{\alpha})$ functions in $R^{(i)}_{rst}$ by
combination of
$\hat{\delta}^{-2}(X^{-\dot{a}}_{\alpha})\hat{\delta}^{+2}(X^{+a}_{\alpha})$.

\section{Kinematical solutions}
This chapter is based on \cite{Triangle_coefficients}. For the case
of triple cuts for the cut momenta $l_i^{\alpha\dot{\alpha}}$ and
associated with them spinors one can find explicit expressions in
terms of external momenta data and t parameter which is the
remaining of the loop integration. First one can define massless
projections of combinations of external momenta:
\begin{equation}
K_1^{\flat\alpha\dot{\alpha}}=\frac{K_1^{\alpha\dot{\alpha}}-
K_2^{\alpha\dot{\alpha}}K_1^2/\gamma}{1-K_1^2K_2^2/\gamma^2},~
K_2^{\flat\alpha\dot{\alpha}}=\frac{K_2^{\alpha\dot{\alpha}}-
K_1^{\alpha\dot{\alpha}}K_2^2/\gamma}{1-K_1^2K_2^2/\gamma^2}.
\end{equation}
Here $\gamma=(K_1,K_2)\pm((K_1,K_2)^2-K_1^2K_2^2)^{1/2}$ which
corresponds to two possible kinematical solutions $\pm S$. In
general $K_j^2\neq0,~j=1...3$.

Using these massless projections one can define corresponding
spinors. The notations of \cite{Triangle_coefficients}
$\lambda_{K_i^{\flat}}^{\alpha}=\langle K_i^{\flat}|\equiv\langle
K_i^{\flat-}|$ and
$\tilde{\lambda}_{K_i^{\flat}}^{\dot{\alpha}}=[K_i^{\flat}|\equiv\langle
K_i^{\flat+}|$ are used in the sense that:
$$
\langle ij\rangle=\langle
i^-|j^+\rangle=\overline{u}_{-}(k_i)u_{+}(k_j),~[ij]=\langle
i^+|j^-\rangle=\overline{u}_{+}(k_i)u_{-}(k_j),
$$
where $u_{\pm}$ are four component Weyl spinors. Then we have:
\begin{equation}
\langle l^-_i|=t\langle K_1^{\flat-}|+\alpha_{i1}\langle
K_2^{\flat-}|,~\langle l^+_i|=\frac{\alpha_{i2}}{t}\langle
K_1^{\flat+}|+\alpha_{i1}\langle K_2^{\flat+}|,
\end{equation}
while for
$l^{\alpha\dot{\alpha}}_i=\sigma_{\mu}^{\alpha\dot{\alpha}}l_i^{\mu}$
one can write:
\begin{equation}
l^{\mu}_i=\alpha_{i2}K^{\flat\mu}_1+\alpha_{i1}K_2^{\flat\mu}+\frac{t}{2}\langle
K_1^{\flat-}|\gamma^{\mu}|K_2^{\flat-}\rangle+\frac{\alpha_{i1}\alpha_{i2}}{2t}\langle
K_2^{\flat-}|\gamma^{\mu}|K_1^{\flat-}\rangle.
\end{equation}
Here the explicit expressions in terms of $K_i^2$ and $\gamma$ for
$\alpha_{ij}$ can be found in appendix A of
\cite{Triangle_coefficients}:
\begin{eqnarray}
\alpha_{01}&=&\frac{K_1^2(\gamma-K_2^2)}{\gamma^2-K_1^2K_2^2},
~\alpha_{02}=\frac{K_2^2(\gamma-K_1^2)}{\gamma^2-K_1^2K_2^2},\nonumber\\
\alpha_{11}&=&\alpha_{01}-\frac{K_1^2}{\gamma},~\alpha_{12}=\alpha_{02}-1,\nonumber\\
\alpha_{21}&=&\alpha_{01}-1,~\alpha_{22}=\alpha_{02}-\frac{K_2^2}{\gamma}.
\end{eqnarray}

Using these expressions one has the following relations for the
spinor products of $\langle l_il_j \rangle$:
\begin{eqnarray}
\langle l_1l_2\rangle&=&-t\left(1-\frac{K_1^2}{\gamma}\right)\langle
K_1^{\flat}K_2^{\flat}\rangle,\nonumber\\
\langle l_1l_3\rangle&=&\frac{tK_1^2}{\gamma}\langle
K_1^{\flat}K_2^{\flat}\rangle,\nonumber\\
\langle l_2l_3\rangle&=&t\langle K_1^{\flat}K_2^{\flat}\rangle.
\end{eqnarray}

In the case when $K_j^2=0$ fore some $j$ ($T^{2m}$ scalar integral)
the explicit solution is different. If the massless leg is attached
to the $\overline{\mbox{MHV}}$ vertex then the solution takes the
form (here we assume that $K_2^2=0$):
\begin{eqnarray}
\langle l_3^-|&=&\frac{t}{[K_2K_1^{\flat}]}\langle
K_2^+|K_1+\frac{K_1^2}{\gamma}\langle\chi^-|,~\langle
l_3^+|=\frac{[\chi K_1^{\flat}]}{[K_2 K_1^{\flat}]}\langle
K_2^+|,\nonumber\\
\langle l_1^-|&=&\frac{1}{[K_2 K_1^{\flat}]}\langle
K_2^+|K_1,~\langle
l_1^+|=-t\frac{[K_1^{\flat}\chi]}{[K_1K_1^{\flat}]}\langle
K_2^+|-\langle K_1^{\flat}|\nonumber\\
\langle l_2^-|&=&\frac{1}{[\chi K_1^{\flat}]}\langle
K_1^{\flat}|K_3+\frac{t}{[K_2 K_1^{\flat}]}\langle
K_2^{+}|K_1,~\langle l_2^+|=\langle l_3^+|,
\end{eqnarray}
where $\langle\chi|$ is arbitrary spinor. Using these solutions we
have the following relevant for our computations results:
\begin{eqnarray}
\langle l_1l_2\rangle&=&-\frac{K_3^2}{\gamma}\langle\chi K_1^{\flat}\rangle,\nonumber\\
\langle l_1l_3\rangle&=&-\frac{K_1^2}{\gamma}\langle\chi K_1^{\flat}\rangle,\nonumber\\
\langle l_2l_3\rangle&=&\frac{[K_1^{\flat}K_2]}{[\chi
K_1^{\flat}]}\langle l_3K_2 \rangle.
\end{eqnarray}

\section{Scalar integrals}
For the Box type integral:
\begin{equation}
B_{K_1^2,K_2^2,K_3^2,K_4^2}=\int
\frac{d^{D}l}{(2\pi)^{D}}\frac{1}{l^2(K_1+l)^2(K_1+K_2+l)^2(l-K_4)^2},
\end{equation}
where $D=4-2\epsilon$ we define dimensionless function
$\mathcal{B}_{K_1^2,K_2^2,K_3^2,K_4^2}$ as:
\begin{equation}
\mathcal{B}_{K_1^2,K_2^2,K_3^2,K_4^2}=\left(
i\pi^{D/2}r_{\Gamma}\right)^{-1}\Delta(2\pi)^{D}B_{K_1^2,K_2^2,K_3^2,K_4^2},
\end{equation}
where
$$
r_{\Gamma}=\frac{\Gamma(1-\epsilon)\Gamma(1+\epsilon)}{\Gamma(1-2\epsilon)},
$$
while for $\Delta$ for $B^{3m}$, $B^{2mh}$, $B^{2me}$ and $B^{1m}$
box scalar integrals we have ($s_{ij}=(K_i+K_j)^2$)
\begin{eqnarray}
\Delta^{3m}&=&s_{12}s_{23}-K_2^2K_4^2,\nonumber\\
\Delta^{2mh}&=&s_{12}s_{23},\nonumber\\
\Delta^{2me}&=&s_{12}s_{23}-K_2^2K_4^2,\nonumber\\
\Delta^{1m}&=&s_{12}s_{23}.\nonumber\\
\end{eqnarray}
For the $\mathcal{B}$ functions we have \cite{QCD_One_loop_Int}
(note that we rearranged IR divergent part of $B^{2mh}$ in
comparison with \cite{QCD_One_loop_Int}):
\begin{eqnarray}
\mathcal{B}^{2mh}(1,2|K_3^2,K_4^2)&=&\frac{\mu^{-2\epsilon}}{\epsilon^2}\left(s_{12}^{\epsilon}+2s_{23}^{\epsilon}-K_3^{2\epsilon}-K_4^{2\epsilon}\right)
+\mbox{log}\left(\frac{K_3^2}{s_{12}}\right)\mbox{log}\left(\frac{K_4^2}{s_{12}}\right)-\nonumber\\
&&-2\mbox{Li}_2\left(1-\frac{K_3^2}{s_{23}}\right)-2\mbox{Li}_2\left(1-\frac{K_4^2}{s_{23}}\right)-\mbox{Log}^2\left(\frac{s_{12}}{s_{23}}\right)+O(\epsilon),\nonumber\\
\mathcal{B}^{2me}(1,3|K_2^2,K_4^2)&=&\frac{\mu^{-2\epsilon}}{\epsilon^2}\left(2s_{12}^{\epsilon}+2s_{23}^{2\epsilon}-2K_2^{2\epsilon}-2K_4^{2\epsilon}\right)
-2\mbox{Li}_2\left(1-\frac{K_2^2}{s_{13}}\right)-\nonumber\\
&&-2\mbox{Li}_2\left(1-\frac{K_2^2}{s_{23}}\right)-2\mbox{Li}_2\left(1-\frac{K_4^2}{s_{12}}\right)-2\mbox{Li}_2\left(1-\frac{K_4^2}{s_{23}}\right)\nonumber\\
&&+2\mbox{Li}_2\left(1-\frac{K_2^2K_4^2}{s_{12}s_{23}}\right)-\mbox{Log}^2\left(\frac{s_{12}}{s_{23}}\right)+O(\epsilon),\nonumber\\
\mathcal{B}^{1m}(1,2,3|K_4^2)&=&\frac{\mu^{-2\epsilon}}{\epsilon^2}\left(2s_{12}^{\epsilon}+2s_{23}^{\epsilon}-2K_4^{2\epsilon}\right)
-2\mbox{Li}_2\left(1-\frac{K_4^2}{s_{12}}\right)-2\mbox{Li}_2\left(1-\frac{K_4^2}{s_{23}}\right)-\nonumber\\
&&-\mbox{Log}^2\left(\frac{s_{12}}{s_{23}}\right)-\frac{\pi^2}{3}+O(\epsilon).
\end{eqnarray}

For triangle scalar integrals
\begin{equation}
T_{K_1^2,K_2^2,K_3^2}=\int \frac{d^{D
}l}{(2\pi)^{D}}\frac{1}{l^2(K_1+l)^2(l-K_3)^2},
\end{equation}
we have similar definitions:
\begin{equation}
\mathcal{T}_{K_1^2,K_2^2,K_3^2}=\left(
i\pi^{D/2}r_{\Gamma}\right)^{-1}\Delta(2\pi)^{D}T_{K_1^2,K_2^2,K_3^2}.
\end{equation}
While for $\Delta$ coefficients for the relevant for our discussion
cases ($T^{2m}$ and $T^{3m}$ scalar triangles with $q^2$ massive
leg) we have:
\begin{eqnarray}
\Delta^{3m}&=&q^2,\nonumber\\
\Delta^{2m}&=&K_2^2-q^2.
\end{eqnarray}
\begin{eqnarray}
\mathcal{T}^{2m}(1|K_2^2,q^2)&=&\frac{\mu^{-2\epsilon}}{2\epsilon^2}\left(K_2^{2\epsilon}-q^{2\epsilon}\right).
\end{eqnarray}
The $T^{3m}$ triangle is IR finite and the answer for it is given in
terms of Davydychev function \cite{Usyukina:1992jd}
$\mathcal{T}^{3m}(K_1^2,K_2^2,q^2)=\mathcal{T}^{3m}(K_1^2/q^2,K_2^2/q^2)$:
\begin{eqnarray}
\mathcal{T}^{3m}&=&\frac{2\mbox{Li}_2(-xR)+2\mbox{Li}_2(-yR)+\mbox{Log}(xR)\mbox{Log}(yR)+\mbox{Log}\left(\frac{y}{x}\right)\mbox{Log}\left(\frac{1+yR}{1+xR}\right)}{Q}\nonumber\\&+&\frac{
\pi^2}{3Q},\nonumber\\
Q&=&\left((1-x-y)^2-4xy\right)^{1/2},~R=2(1-x-y+R)^{-1},~x=\frac{K_1^2}{q^2},~y=\frac{K_2^2}{q^2}.\nonumber\\
\end{eqnarray}

\newpage
 \begin{figure}[h]
 \begin{center}
 \leavevmode
  \epsfxsize=8cm
 \epsffile{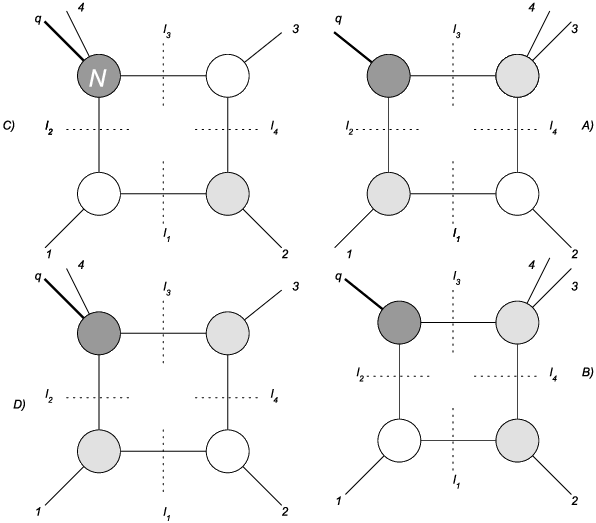}
\epsfxsize=9cm
 \epsffile{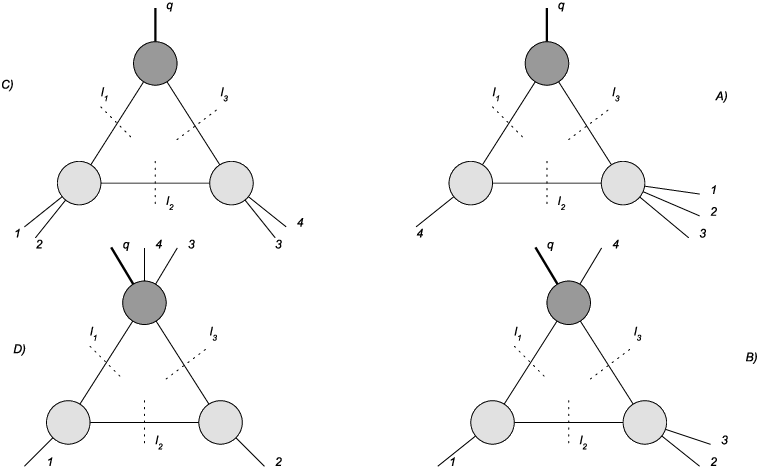}
\epsfxsize=9cm
 \epsffile{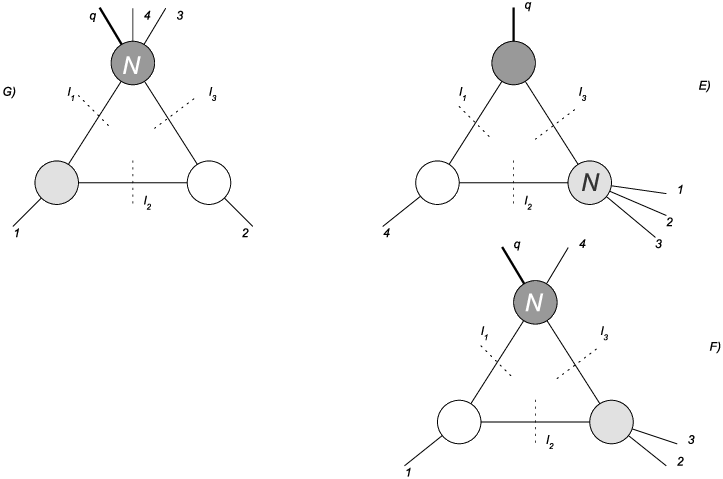}
 \end{center}\vspace{-0.2cm}
 \caption{List of contributing quadruple and all possible triple cuts
 for NMHV four point form factor. B), G), D), F) triple cuts give vanishing
 results for corresponding coefficients. Permutations of external momenta are not shown.}\label{fig_NMHV4_quadr_triple_cuts}
 \end{figure}

\end{document}